\documentclass[letterpaper, 10 pt, conference]{ieeeconf}  

\IEEEoverridecommandlockouts                              

\usepackage{comment}
\usepackage{graphics} 
\usepackage{epsfig} 
\usepackage{times} 
\usepackage{amsmath} 
\usepackage{amssymb}  
\usepackage{bbold}  
\usepackage[space]{cite}
\usepackage{xcolor}
\usepackage{enumerate}
\newtheorem{remark}{Remark}

\newtheorem{proposition}{Proposition}

\title
{\LARGE \bf Transferring the driveshaft inertia to the grid via the DC-link in MV drive systems}

\author{Catalin Arghir$^{1}$,  Pieder Jörg$^{2}$,  Silvia Mastellone$^{1}$
\thanks{This work was supported by the Swiss National Science Foundation under NCCR Automation, grant agreement 51NF40\_180545}
\thanks{$^{1}$C. Arghir and S. Mastellone are with the Institute of Electric Power Systems, University of Applied Sciences Northwest Switzerland, Windisch, Switzerland. \texttt{carghir@ethz.ch, silvia.mastellone@fhnw.ch}}
\thanks{$^{2}$P. Jörg is with ABB Motion, ABB Switzerland Ltd., Turgi, Switzerland.  \texttt{pieder.joerg@ch.abb.com}}
}

\begin{document}

\maketitle
\thispagestyle{empty}
\pagestyle{empty}

\begin{abstract}


This paper investigates a control approach that renders the total driveshaft inertia available to the grid while enhancing the fault ride-through (FRT) capability of medium-voltage (MV) drive systems. Two main contributions are presented. First, we show how the rotational inertia of the driveline can be synchronously coupled to the grid through modifications to the speed-control reference and the DC-link control strategy. For the latter, we investigate two alternatives: conventional cascaded control and synchronous-machine (SM) model matching. Second, we show that both the standard phase-locked loop (PLL) and the matching control approach can be interpreted, via the ray–circle complementarity, as gradient-based control schemes. This perspective provides new insight into matching control by revealing its embedded PLL dynamics and clarifying its current-limiting and reference-tracking properties. Finally, an extensive simulation study evaluates the proposed approaches and their FRT performance.

\end{abstract}

\section{INTRODUCTION}

Medium-voltage (MV) drive systems play a central role in high-power industrial processes, electric propulsion, renewable and conventional energy generation \cite{kouro2012powering}. 
Such processes are grid-connected through converters supplying a complex driveline through one or more electric machines. 
%
%
%
When equipped with an active front end—particularly in the high-power MV range—such drives are increasingly expected to contribute beyond basic specifications, by providing ancillary services such as inertia, phase jump support, and reactive power compensation \cite{gomis2022grid}. Ubiquitous in wind applications, emerging standards such as IEEE 1547.3 and IEC 61400 are becoming relevant to large industrial drives \cite{hokayem2024control}, requiring them to support the natural behavior of the synchronous machine \cite{9782556, mourouvin2025direct}. 

In addition, established regulations (e.g., \cite{Netz}) and ongoing standardization efforts are promoting the transition of large electricity consumers from passive loads to active grid-supporting assets by leveraging the capabilities of power-electronic converters already embedded in modern industrial and mobility infrastructures. One example is the European working group developing a standard on the self-regulation of dispatchable loads (see \cite{CENELAC}), based on IEC TS 62898-3-3. The standard addresses frequency and voltage stabilization in AC networks through loads that autonomously adjust their active power consumption in response to grid conditions.

Traditionally, ancillary services such as voltage control, frequency stabilization, and power quality management have been provided mainly by centralized generation units or dedicated grid equipment. However, the increasing electrification of industry, transport, and heating, together with the growing penetration of renewable energy sources, has significantly changed grid dynamics. Converter-driven loads are now widespread, fast-acting, and highly controllable, making them a promising resource for grid support. Consequently, there is strong interest in equipping converters controlling large loads with strategies that enable reliable ancillary service provision without compromising their primary industrial or commercial functions.

While significant attention has been paid to motor-side control \cite{geyer2016model} 
and grid-side compliance \cite{shakerighadi2023overview, stanojev2025grid, he2024cross}, less is understood about how mechanical energy stored in the drivetrain can interact dynamically with the grid \cite{liu2024comparative, 6900057}, for example, in cases where torsional resonances overlap with sub-synchronous harmonics. In this work, we adopt an interconnected viewpoint and show how, whether acting as generators or loads, MV drive systems can play a significant role in grid stabilization.

{
Unlike conventional grid-forming converters, which emulate inertia through a virtual synchronous machine model or synthetic inertia loop \cite{8506338, mallemaci2021comprehensive}, the approach proposed here exploits the physical rotational inertia already present in the drivetrain. Through a modification of the DC-link control structure, this mechanical inertia becomes dynamically visible at the grid interface. In this sense, the proposed method does not introduce virtual inertia but rather transfers existing mechanical inertia to the electrical domain.
}
%

We build on the control-theoretic foundations of \cite{arghir2019transverse}, which provides stability guarantees for both single- and multi-inverter settings. A key feature of this framework is the inherent passivity induced by matching control, ensuring a robust grid interaction \cite{10598346}. Furthermore, we highlight its connection to the Online Feedback Optimization (OFO) paradigm \cite{HAUSWIRTH2024100941}, which points to potential future research. To enable a fair comparison, we propose a modification to the reference signals in the classical cascaded-control structure. By using the same pumped-hydro drive setup as in \cite{chandrasekaran2025reactive}, we further expand in the direction of drive availability, and provide a benchmark scenario for the matching approach.

Our work is thus distinct in combining: ($i$) a first-principles treatment of synchronization mechanisms (PLL and matching control) as gradient-based algorithms; ($ii$) a demonstration of natural inertia transfer by physical energy coupling while satisfying current limits, tracking performance, and providing voltage support; ($iii$){ a comparative study  examining converter operation under both nominal and faulty grid conditions using a high-fidelity simulation of an industrial drive system. This study enables validation of the effectiveness of the proposed method and facilitates the assessment of its operational limits under realistic grid disturbances. }


The remainder of this paper is structured as follows. Section \ref{prbState} presents the modeling framework and coupling control formulation. Section \ref{ctrlPrb} revisits the standard PLL and matching control. Section \ref{expRes} presents the experimental results, and Section \ref{Concl} concludes the paper with a discussion of implications for future drive system control design.


\section{Drive system control}
\label{prbState}

We consider a back-to-back drive system consisting of a grid-side converter, a DC-link capacitor, a motor-side converter, an electrical machine, and a flexible driveshaft attached to a prime mover or a load. The main role of the grid-side converter, is to deliver active power by regulating the DC-link voltage, and thus allow the motor-side converter to supply the load under a wide range of external conditions.

\subsection{Drive system model}

We adopt an energy-preserving, average-switch model of a back-to-back drive system with a 2-mass driveshaft 
\begin{subequations}
\label{system_eq}
\begin{align}
    L_g\dot{i}_g &= - R_g i_g  + v_g - m_g v_{dc} \label{igdynamics}
    \\
    C_{dc}\dot{v}_{dc} &=  m_g^\top i_g - \tfrac{{\tau}_m}{v_{dc}}{w}_1 
    \label{vdcdynamics}
    \\
    \dot{x}_1 &= w_1
    \\
    M_1\dot{w}_1 &= -C_{12}(w_1-w_2) -K_{12}(x_1-x_2) + {\tau}_m  \label{omega2dynamics} 
    \\
    \dot{x}_2 &= w_2
    \\
    M_2\dot{w}_2 &= -C_{12}(w_2-w_1) -K_{12}(x_2-x_1) - \tau_l \,, \label{omega1dynamics}
\end{align}
\end{subequations}
where $M = M_1\!+\!M_2$ is the total moment of inertia of the driveshaft, $C_{12}$ is the damping coefficient and $K_{12}$ is the spring coefficient of the coupling between the two masses, $w_1, w_2$ are the angular velocities of the two shaft sections, $x_1, x_2$ their absolute angles, ${\tau}_m$ the motor torque reference (control input), and $\tau_l$ is the driveshaft load torque (unmeasured disturbance). Note that this model exhibits a torsional natural frequency (TNF), and therefore the drive requires active damping. 

The DC-link voltage $v_{dc}$ is modeled across a capacitance $C_{dc}$. 
Furthermore, $m_g\in\mathbb{R}^2$ is the grid-side converter $\alpha\beta$-modulation vector (control input), $L_g,{R}_g>0$ phase inductance and resistance, respectively, $i_g\in\mathbb{R}^2$ is the grid-side converter current and $v_g\in\mathbb{R}^2$ is the point-of-common-coupling (PCC) voltage in $\alpha\beta$-coordinates (measured disturbance). We use the well established notions of instantaneous power \cite{o2019geometric} and the power-invariant $\alpha\beta$-frame transformation
\begin{align}
    x_{abc} \mapsto x_{dq0} &= {T}_{\theta} T_{\alpha\beta\gamma} x_{abc} = {T}_{\theta}  x_{\alpha\beta\gamma} \,,
\end{align}
with $T_{\alpha\beta\gamma}^{-1} = T_{\alpha\beta\gamma}^\top$, and where ${T}_{\theta} = \mathrm{blkdiag}({R}_\theta, 1)$, with ${R}_\theta = \begin{bmatrix}\begin{smallmatrix} \cos\theta & -\sin\theta \\ \sin\theta & \cos\theta\end{smallmatrix}\end{bmatrix}$ being the rotation matrix.  Moreover, by virtue of the third-harmonic injection, the maximum modulation amplitude in $\alpha\beta$-coordinates becomes $\tfrac{1}{\sqrt{2}}$. In the rest of this paper we only consider the first two components, namely the two-dimensional vector $x_{dq}$ or $x_{\alpha\beta}$. We are therefore able to define the (instantaneous) active and reactive power for quantities in either $dq$ or $\alpha\beta$ as
\begin{align}
    P &= v_{\alpha\beta}^\top i_{\alpha\beta} = v_{dq}^\top i_{dq}
    \\
    Q &= v_{\alpha\beta}^\top Ji_{\alpha\beta} = v_{dq}^\top Ji_{dq} \,,
\end{align}
where $J = {R}_{\pi/2}$, and where the $dq$-frame is attached to the grid angle. As we shall see, we only need the $dq$ transformation for the integral term of the proportional-integral (PI) current control to ensure tracking of a harmonic reference, and henceforth all equations can be equivalently expressed in either coordinates.
\begin{remark}\label{modelassumption}
    In this model, we omit the motor and its torque control dynamics, an abstraction which enables us to focus on the grid-side behavior and its interaction with the shaft dynamics. 
\end{remark}

\begin{remark}\label{modelchoice}
Our particular choice of modeling assumptions stems from the fact that \eqref{system_eq} is rendered passive with input $(v_g,\tau_l)$, output $(i_g,w_2)$ and storage function 
\begin{align}\label{energy1}
    {H} &= \tfrac{1}{2} {i}_g^\top L_g {i}_g  + \tfrac{1}{2} C_{dc}{v}_{dc}^2 + \tfrac{1}{2} M_{1}{w}_{1}^2 + \tfrac{1}{2} M_{2}{w}_{2}^2  \notag
    \\ 
    &+ \tfrac{1}{2}K_{12}(x_1-x_2)^2\,.
\end{align}
Due to the skew-symmetry of the system equations, it is easy to check that the passivity conditions \cite{hill1976stability} are satisfied.
\end{remark}

\subsection{Drive system objectives}
\label{control_section} 
 
On a high level, the driveshaft is required to rotate at a given speed while the grid-side converter supplies the motor-side converter with a stable voltage. The tracking objectives
\begin{equation}
\label{trackingobjectives}
\begin{split}
    {w}_1, w_2 &\rightarrow {w}^\mathrm{ref}  
    \\
    v_{dc} &\rightarrow v_{dc}^\mathrm{ref} 
\end{split}
\end{equation}
are meant to reject the external disturbance $(\tau_l, v_g)$. Moreover, $|{w}^\mathrm{ref}|\leq w_\mathrm{nom}$ is an external set-point, while $|\tau_l|\leq \tau_\mathrm{nom}$, $\|v_g\|\leq v_{g,\mathrm{nom}} = \alpha_v v_{g,\mathrm{nom}}$ with e.g. $\alpha_v>1$ representing a grid code overvoltage factor. We also consider some margin for producing reactive power necessary for the loose regulation of the PCC voltage magnitude $v_g$. 
\begin{align}\label{maxpower}
    \tau_\mathrm{nom}w_\mathrm{nom} = P_{g,\mathrm{nom}} = \tfrac{1}{\alpha_q} \sqrt{P_{g,\mathrm{nom}}^2 + Q_{g,\mathrm{nom}}^2} \,,
\end{align}
where e.g. $\alpha_q>1$ is a de-rating factor, such that the maximum current amplitude becomes $i_{g,\mathrm{nom}} = \tfrac{\alpha_q P_{g,\mathrm{nom}}}{v_{g,\mathrm{nom}}}$. 
%

In a classical back-to-back drive system, the control strategy is a two-fold tracking cascade, one on the load side: a speed regulator provides reference to the inner torque regulator, and one on the grid-side: a DC-link regulator provides reference to the inner current regulator. Both inner loops determine the modulation signal to the corresponding converter. Starting from this structure, and omitting from our analysis the inner torque control loop, we shall provide a methodology that makes the entire drive-shaft appear as {\it synchronous inertia}\footnote{i.e. the inertia of a synchronous machine directly connected to the PCC} at the grid-side. Our approach consists of two steps, the first step is to use the speed control to elastically couple the driveshaft to the DC-link capacitor interpreted as a rotational mass. The second step addresses the grid-side converter control.

\subsection{Speed control and coupling}

In our first step, we exploit the way the motor torque couples $M_1$ and $C_{dc}$ by implementing a virtual spring-damper element. To do this, we recast the speed controller by proposing a coordinate transformation that makes $v_{dc}$ appear as the angular velocity of an additional mass element to be coupled to the drive-shaft
\begin{align}\label{Mdc}
    v_{dc} \mapsto w_{dc} = \tfrac{w^\mathrm{ref}}{v_{dc}^\mathrm{ref}}v_{dc} \,.
\end{align}
We then replace the reference $w^\mathrm{ref}$ by $w_{dc}$ in the standard PI speed control loop
\begin{subequations}
\label{speedcontrol_eq}
\begin{align}\label{speed_PI}
    \dot{x}_m &=  {w}_1-{w}_\mathrm{dc} 
    \\
    \tau_m &= \underset{|\cdot| \leq \tau_\mathrm{nom}}{\mathrm{sat}} -K_{p,m}({w}_1-{w}_{dc}) - K_{i,m}x_m\,,
\end{align}
\end{subequations}
where $K_{p,m}$, and $K_{i,m}$ are control gains, and where $\mathrm{sat}$ denotes the presence of integrator anti-windup. 
We shall now see the effects of the coordinate transformation \eqref{Mdc} and controller \eqref{speedcontrol_eq} on system \eqref{system_eq}.
\begin{proposition}\label{passivity1}
Assume that the state-space $\mathcal{X}$ admits only nonzero values for $w_{dc}$. Define the resulting mass-spring-damper elements as
\begin{subequations}
\label{param_Mdc}
\begin{align}
    M_{dc} &= \big(\tfrac{v_{dc}^\mathrm{ref}}{w^\mathrm{ref}}\big)^2 C_{dc}
    \\
    K_{01} &= \big(\tfrac{w_1}{w_{dc}}\big) K_{i,m}
    \\
    C_{01} &= \big(\tfrac{w_1}{w_{dc}}\big) K_{p,m} \,,
\end{align}
\end{subequations}
and the angle of the mass element associated with the DC-bus as $x_0 = x_1 - x_m$. Consider that the modulation signal $m_g$ is a to-be-defined feedback law. Then, system \eqref{system_eq}, under the coordinate and feedback transformation \eqref{Mdc}-\eqref{speedcontrol_eq}, is passive with input $(v_g,\tau_l)$, output $(i_g,w_2)$ and storage function 
\begin{align}\label{energy2}
    \mathcal{H} &= \tfrac{1}{2} {i}_g^\top L_g {i}_g  + \tfrac{1}{2} M_{dc}{w}_{dc}^2 + \tfrac{1}{2} M_{1}{w}_{1}^2 + \tfrac{1}{2} M_{2}{w}_{2}^2  \notag
    \\ 
    &+ \tfrac{1}{2}K_{12}(x_1-x_2)^2\,. 
\end{align}
\end{proposition}
\begin{proof}
System \eqref{system_eq} with feedback \eqref{speedcontrol_eq} becomes
\begin{subequations}
\label{system_Mdc}
\begin{align}
    L_g\dot{i}_g =& - R_g i_g  + v_g - \big(\tfrac{v_{dc}^\mathrm{ref}}{w^\mathrm{ref}}\big) m_g w_{dc} \,, \label{igdynamics_c}
    \\
    \dot{x}_{0} =&~ w_{dc}
    \\
    M_{dc}\dot{{w}_{dc}} =& -C_{01}(w_{dc}-w_1)-K_{01}(x_{0}-x_1)    \notag 
    \\
    &+\big(\tfrac{v_{dc}^\mathrm{ref}}{w^\mathrm{ref}}\big) m_g^\top i_g \label{omega0dynamics}
    \\
    \dot{x}_1 =&~ w_1
    \\
    M_1\dot{w}_1 =& -\tfrac{w_{dc}}{w_1}C_{01}(w_{1}-w_{dc}) -\tfrac{w_{dc}}{w_1}K_{01}(x_{1}-x_{0}) \notag
    \\
    &-C_{12}(w_1-w_2) -K_{12}(x_1-x_2) \label{omega1dynamics_c}
    \\
    \dot{x}_2 =&~ w_2
    \\
    M_2\dot{w}_2 =& -C_{12}(w_2-w_1) -K_{12}(x_2-x_1) - \tau_l \,.\label{omega2dynamics_c}
\end{align}
\end{subequations}
To check that the passivity conditions hold, we evaluate
\begin{align}\label{energydot}
    \dot{\mathcal{H}} &= -{i}_g^\top R_g {i}_g  - C_{12}({w}_{1} - {w}_{2})^2 + i_g^\top v_g + w_2 \tau_l \,.
\end{align}

\end{proof}

\begin{figure}[!ht]
\centering{
\includegraphics[width=0.9\columnwidth]{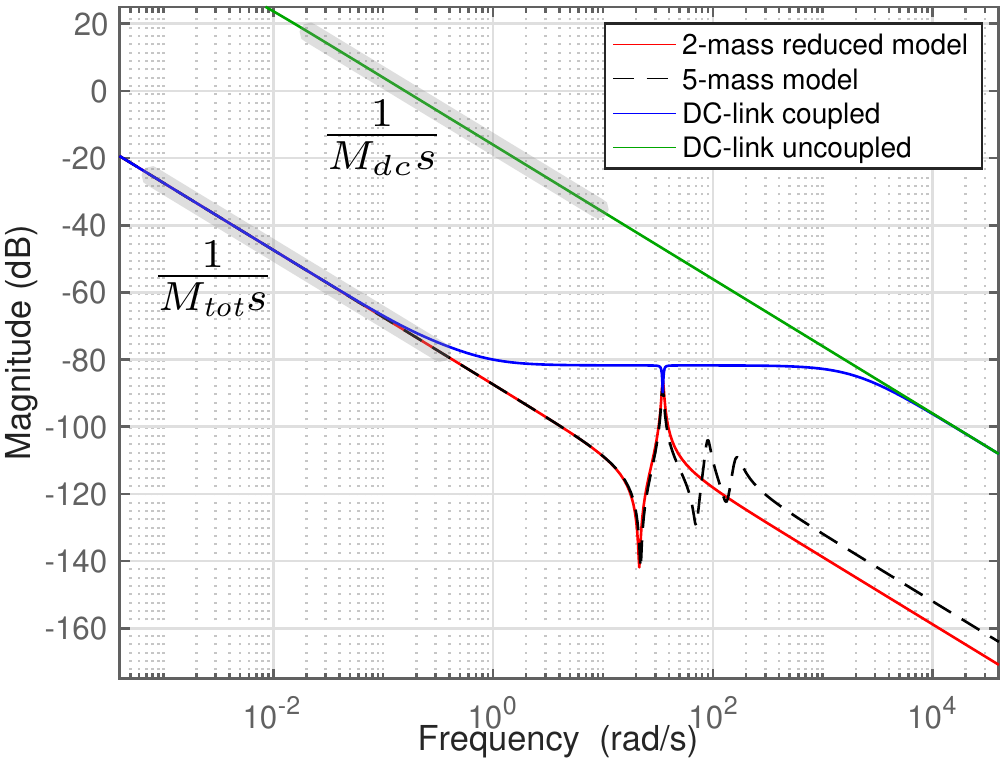}
\caption{Bode plot showing the DC-link impedance compared to the driveshaft inertia as transfer function from torque to speed or, equivalently, input current to voltage. The main driveline torsional resonance, of about $5.5Hz$, appears as a zero in the capacitor dynamics (blue). We see that the coupled system assumes a very large total inertia $M_\textit{tot}$ at low frequency (i.e. within the speed control bandwidth). In this example, the driveshaft inertia is about 3 orders of magnitude higher than that of the DC-link.}	\label{Fig: dcmass}}
\end{figure}

\section{Grid-side control}
\label{ctrlPrb}

The second step in our approach is to induce the behavior of a synchronous machine using the grid-side converter. We achieve this in two ways: ($i$) via traditional cascaded-control by drooping the DC-bus with grid frequency, and ($ii$), by matching the dynamics of a SM by directly acting on the modulation vector.

\begin{figure}[!ht]
\centering{
\includegraphics[width=1\columnwidth]{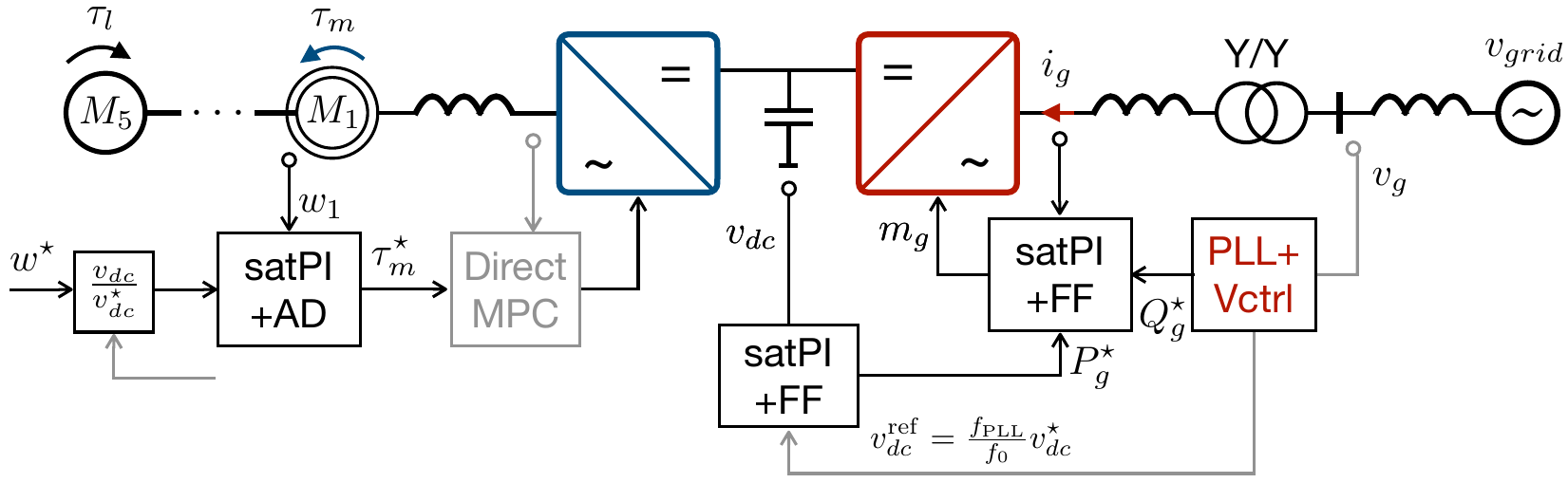}
\caption{Block diagram of the cascaded PI control setup. From left to right: speed control, torque control, voltage control and current control loops typically implemented in a drive system, together with a PLL that extracts the grid voltage and the angle for the $dq$-transformation. The anti-windup tracking controllers are denoted by satPI, while further feed-forward (FF) and active-damping (AD) terms are appropriately added.} 	\label{Fig: cascBlock}}
\end{figure}

As seen in Fig. \ref{Fig: cascBlock}, the PCC voltage $v_g$ is attached to an infinite bus $v_\textit{grid}$ through the grid impedance, not accounted for in our model. When this impedance is large, the short circuit ratio (SCR) is small and the grid is considered weak. In this case, the PCC is weakly regulated and therefore reactive power compensation is crucial. In addition, the PCC is susceptible to noise and local disturbances and therefore a PLL is required to extract a stable frequency and angle.

\subsection{Cascaded PI approach} \label{sec: cascPI}

For our first approach, we modify the DC-bus reference in the cascaded-control structure as
\begin{align}\label{ref2PLL}
    v_{dc}^\star = \tfrac{1}{\eta}\omega_\textit{pll} \,,
\end{align}
where $\eta = \tfrac{\omega_0}{v_{dc}^\mathrm{ref}}$ is the SM matching gain, $\omega_0$ the nominal grid frequency, and $(\theta_\textit{pll}, \omega_\textit{pll})$ is the output of the PLL. The DC-link control is defined as
\begin{subequations}
\label{dccontrol_eq}
\begin{align}\label{vdc_PI}
    \dot{x}_{dc} &= v_{dc}-v_{dc}^\star
    \\
    P_g^\star &= \underset{|\cdot| \leq P_{g,\mathrm{nom}}}{\mathrm{sat}}\!\big(-K_{p,dc}(v_{dc}\!-\!v_{dc}^\star) \!-\! K_{i,dc}x_{dc}\big){v_{dc}^\mathrm{ref}},
\end{align}         
\end{subequations}
where $K_{p,dc}$ and $K_{i,dc}$ are control gains. Notice that the DC-link regulator now sees a much larger capacitor due to the coupling control
\begin{align}\label{Ctot}
    C_{tot} = C_{dc} + \big(\tfrac{w^\mathrm{ref}}{v_{dc}^\mathrm{ref}}\big)^2(M_1+M_2) \,.
\end{align}
If we consider that the inner current loop always achieves its goal, we then have $v_{dc}m_g^\top i_g = P_g^\star$. Under this assumption, we can write the closed-loop DC-dynamics via the coordinate transformation $\omega = \eta v_{dc}$ as
\begin{equation} \label{vdcPLL}
    \tfrac{C_{tot}}{\eta^2}\dot\omega = -\tfrac{K_{p,dc}}{\eta^2}(\omega-\omega_\textit{pll}) -\tfrac{K_{i,dc}}{\eta^2}(\theta-\theta_\textit{pll}) - \tfrac{P_m}{\omega} \,,
\end{equation}
where $P_m = \tau_m w_1$ and $\theta = \eta x_{dc} + \theta_\textit{pll}$. In essence, this can be seen as a coupling from the grid frequency to DC-bus voltage. For the grid-side current reference, we consider the circular current limiter defined below
\begin{equation} 
   i_{g}^\star = \underset{\|\cdot\| \leq i_{g,\mathrm{nom}}}{\mathrm{sat}}\tfrac{1}{\|v_g\|^2}\begin{bmatrix}v_g^\top \\ v_g^\top J\end{bmatrix}\begin{bmatrix}P_g^\star \\ Q_g^\star\end{bmatrix} \label{ig_limiter} \,.
\end{equation}

The current controller typically uses the PLL angle to implement the integral term in $dq$-frame 
\begin{subequations}
\label{currentcontrol_eq}
\begin{align}\label{ig_PI}
\dot{x}_{g} &= {R}_{\theta_\textit{pll}}^\top(i_{g}-i_{g}^\star)
    \\
m_g &= \underset{\|\cdot\| \leq \tfrac{1}{\sqrt{2}}}{\mathrm{sat}}\tfrac{1}{v_{dc}}(K_{p,g}(i_{g}-i_{g}^\star) + K_{i,g}{R}_{\theta_\textit{pll}}{x_{g}} \notag
\\
&~~~~~~~~~~~~~+v_g - Z_gi_g^\star) \label{mg_equation} \,,
\end{align} 
\end{subequations}
where $Z_g = \begin{bmatrix}\begin{smallmatrix} R_g & -\omega_0L_g \\ \omega_0L_g & R_g \end{smallmatrix}\end{bmatrix}$ is the converter impedance. Moreover, $K_{p,g}$, and $K_{i,g}$ are control gains. 
Note that the $\mathrm{sat}$ function in \eqref{ig_limiter} and \eqref{mg_equation} represents a circular limiter and, as seen in \cite{chandrasekaran2025reactive}, its implementation is crucial to preserving converter stability. 
\subsection{PLL design}

\begin{figure}[!ht]
\centering{
\includegraphics[width=1\columnwidth]{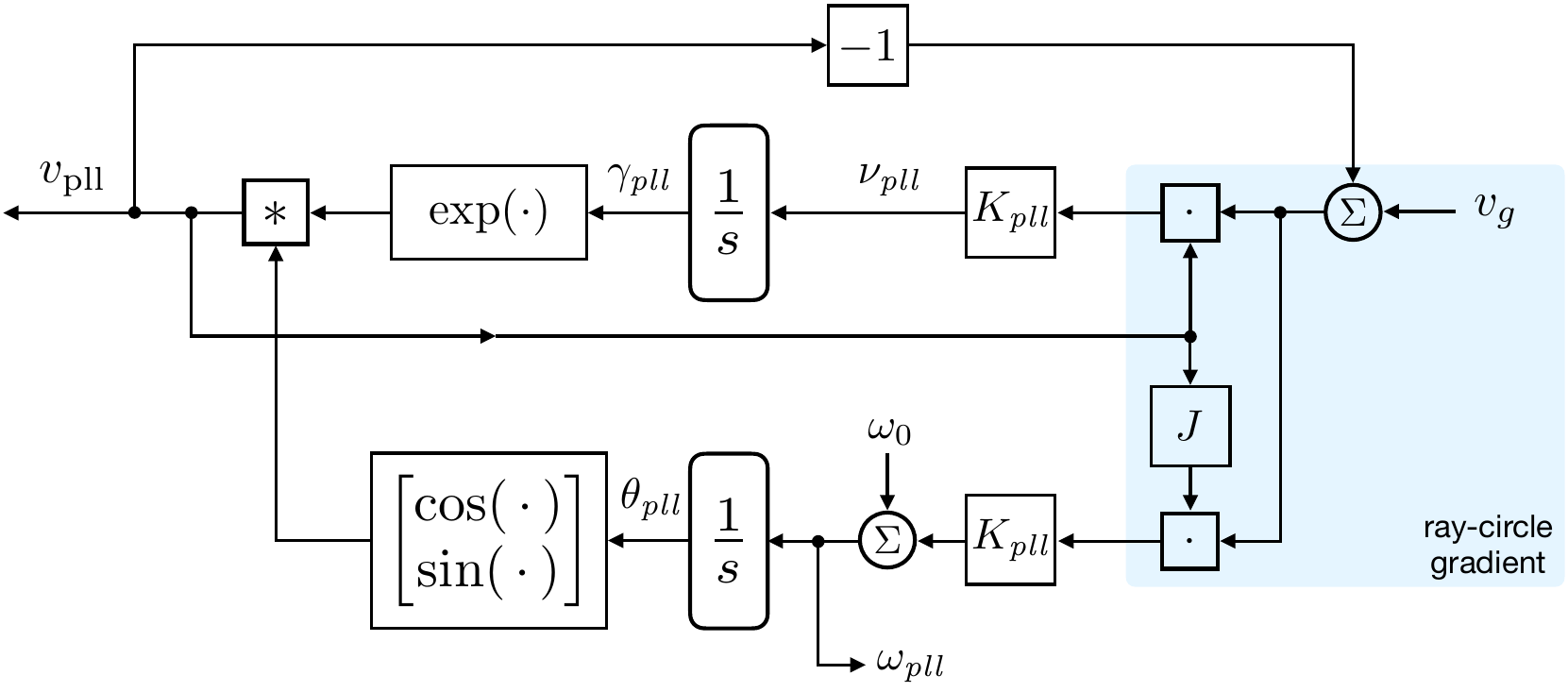}
\caption{Block diagram of the proposed PLL design where the $\cdot$ block represents the dot-product and the $*$ element-wise product.} 	\label{Fig: pllDiag}}
\end{figure}

In this section we propose an alternative PLL design, inspired by the work in \cite{arghir2019transverse}, Section 3.2, where the gradient is decomposed into tangential and transverse components relative to the geometry of interest, that of a circular path in the $\alpha\beta$-frame. We start by defining the energy function as the distance from the measured PCC voltage to the generated PLL output
\begin{align}
    \mathcal{U} = \tfrac{1}{2}(v_\textit{pll} - v_g)^\top(v_\textit{pll} - v_g) \,.
\end{align}
Let us define the PLL output vector $v_\textit{pll}\in\mathbb{R}^2$ via the ray-circle decomposition as
\begin{align}
    v_\textit{pll} = e^{\gamma_\textit{pll}}{R}_{\theta_\textit{pll}}\mathrm{g}_1 \,, 
\end{align}
where $\mathrm{g}_1 = \begin{bmatrix}\begin{smallmatrix} 1 \\ 0 \end{smallmatrix}\end{bmatrix}$ is a unit vector, $\theta_\textit{pll}$ is the PLL angle and where $\gamma_\textit{pll}\in\mathbb{R}$ is a second degree of freedom of the PLL. The advantage of using the {\it log of magnitude} to complement the angle lies in the gradient expression
\begin{align} \label{eq: gradientPLL}
    \nabla_{{\gamma_\textit{pll}}}{v_\textit{pll}} = v_\textit{pll}^\top \,\,,\,\,\,
    \nabla_{\theta_\textit{pll}}{v_\textit{pll}} =  v_\textit{pll}^\top {J}^\top
\end{align}
while $\nabla_{v_\textit{pll}}\mathcal{U} = v_\textit{pll} - v_g$. Therefore, the design of the PLL becomes a gradient descent law via the chain rule
\begin{subequations} \label{eq: ePLL}
\begin{align}
    \dot\gamma_\textit{pll} &= - K_\textit{pll} \nabla_{{\gamma_\textit{pll}}}{v_\textit{pll}}\nabla_{v_\textit{pll}}\mathcal{U} \label{eq: gammaPLL}
    \\
    \dot\theta_\textit{pll} &= - K_\textit{pll} \underbrace{\nabla_{{\theta_\textit{pll}}}{v_\textit{pll}}\nabla_{v_\textit{pll}}\mathcal{U}}_{\nabla\mathcal{U}} +~ \omega_0 \,, \label{eq: thetaPLL}
\end{align}
\end{subequations}
where $K_\textit{pll}$ is a synchronization gain that determines the tracking bandwidth. This translates into the block diagram in Fig. \ref{Fig: pllDiag} and serves as a basis towards the design of the tracking component in the matching control of the next section.
\begin{remark}\label{proofrpll}
Notice that, when $\dot\gamma_\textit{pll}$ is set to zero in \eqref{eq: gammaPLL}, the classical PLL structure emerges (as the gradient simply becomes the sine of angle difference). Our proposed structure is more general, can be used to fully reconstruct the grid voltage, and is in line with recent research \cite{9524491}. One can further check stability by taking $\mathcal{U}$ as Lyapunov function.
\end{remark}

%
%
\begin{remark}\label{coupledpll}
    As we have seen from \eqref{omega0dynamics}, the driveshaft is flexibly coupled to the DC-bus angle, and from \eqref{vdcPLL}, that the DC-bus tracks the PLL angle. Alternatively, the constant feedforward term $\omega_0$ in \eqref{eq: thetaPLL} may be replaced by the matching frequency $\eta v_{dc}$ to yield a bidirectional coupling. We now have a unified mechanism that enables an energy-based interconnection of the entire power conversion chain --- a radically different approach to drive system behavior.
\end{remark}
\begin{remark}\label{rem_vpll}
Note that, in all equations of Section \ref{sec: cascPI}, we may use the output of the PLL $v_\textit{pll}$ instead of the measured PCC voltage $v_g$. This is generally beneficial in weak grid scenarios where this voltage is affected by noise. 
\end{remark}

\subsection{AC voltage control}
As we shall tackle both weak and stiff grid scenarios, we assign the reactive power reference to perform PCC voltage regulation via a feedback of the form
\begin{align} \label{eq: qVC}
    Q_g^\star = \Pi_{\mathcal{C}} K_{p,v} (\|v_{g}\|^2 - v_{g,\mathrm{ref}}^2) \,,
\end{align}
where $v_{g,\mathrm{ref}}\simeq v_{g,\mathrm{nom}}$ is an external reference which, together with $K_{p,v}$, may be chosen according to the grid conditions. Furthermore, the result is projected onto the feasible set $\mathcal{C}$ illustrated 
in \cite{chandrasekaran2025reactive}. This improves the grid-forming ability of our drive by driving a large capacitive current when the grid dips in magnitude, maintaining the active power required by the drive to the extent possible

\subsection{Revisiting the matching control}

\begin{figure}[!ht]
\centering{
\includegraphics[width=1\columnwidth]{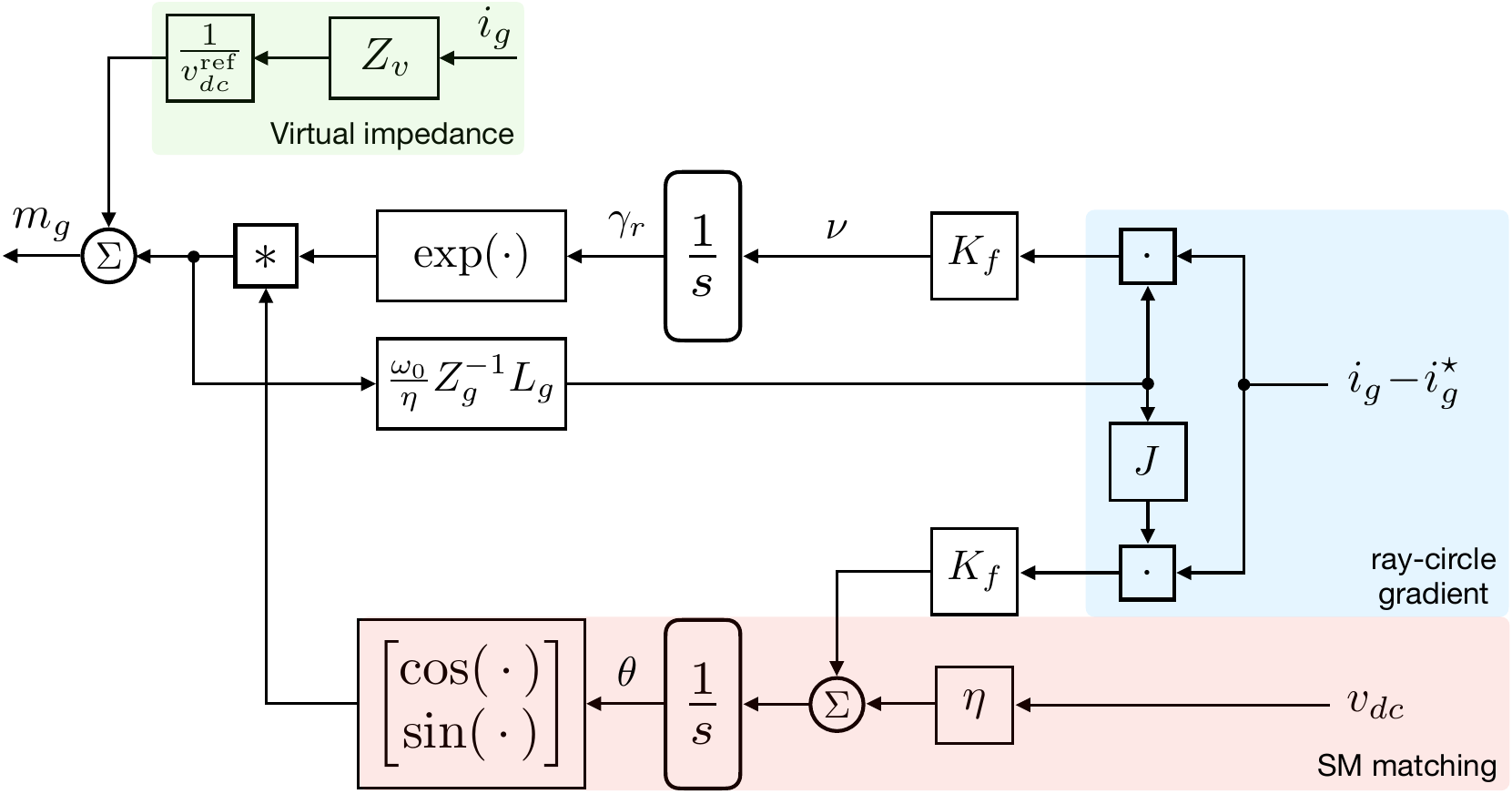}
\caption{Block diagram of the matching control illustrating the SM-matching branch and the ray-circle gradient-based tracking branch.} 	\label{Fig: matchDiag}}
\end{figure}

As we have seen in Fig. \ref{Fig: dcmass}, the DC-bus coupled with the driveline shaft, yields a large DC capacitance within the bandwidth of the speed control loop. We are now in a position to employ the matching control approach \cite{arghir2019electronic, arghir2019transverse} to harness this large DC inertia on the grid-side, while still regulating shaft speed, essentially forming the grid in terms of frequency and voltage. We adopt the PQ-tracking approach developed in \cite{arghir2019transverse}, section 4.3, and augment it with a virtual impedance term. The overall proposal is illustrated in Fig. \ref{Fig: matchDiag}.

In this implementation, we prescribe a PQ setpoint for the PCC, convert it into a current setpoint $i_g^\star$ via \eqref{ig_limiter}, and use it to define an energy function to minimize via modulation angle and log-magnitude. For this purpose, 
\begin{align}
    P_g^\star = \tau_m w_1 
\end{align}
is taken from the speed controller, and $Q_g^\star$ from \eqref{eq: qVC}. We define the synchronization energy function as
\begin{align}
    \mathcal{S} = \tfrac{1}{2}(\hat{i}_g - i_g^\star)^\top L_g(\hat{i}_g - i_g^\star) \,,
\end{align}
where $\hat{i}_g = - Z_g^{-1}\big( \tfrac{1}{\eta}e^{\gamma_r}{R}_\theta\mathrm{g}_1\omega_0 - v_g \big)$ is the steady-state induced by the coordinate $(\gamma_r, \theta)$, as derived in \cite{arghir2018energy,arghir2019transverse}. Notice that, similarly to \eqref{eq: gradientPLL}, we have
\begin{subequations}\begin{align}
    \nabla_{\gamma_r} \hat{i}_g &= -(\tfrac{\omega_0}{\eta} Z_g^{-1}e^{\gamma_r}{R}_\theta\mathrm{g}_1)^\top
    \\
    \nabla_{\theta} \hat{i}_g &= -(\tfrac{\omega_0}{\eta} Z_g^{-1}e^{\gamma_r}{R}_\theta\mathrm{g}_1)^\top J^\top \,.
\end{align}\end{subequations}
To formulate the controller, we take inspiration from the OFO approach in the sense that we replace the steady-state variable $\hat{i}_g$ by the measurement $i_g$ in the gradient of $\mathcal{S}$. The controller dynamics become remarkably similar to the PLL structure \eqref{eq: ePLL}, albeit with a nontrivial steady-state map. 
We express the controller as
\begin{subequations}
\label{matching_law}
\begin{align} 
    m_g &= e^{\gamma_r}{R}_{\theta}\mathrm{g}_1  \label{eq: modlaw}
    \\
    \dot\gamma_r &= - K_f \nabla_{\gamma_r}{\hat{i}_g} L_g({i}_g - i_g^\star) \label{eq: gammaint}
    \\
    \dot\theta &= - K_f  \underbrace{\nabla_{\theta}{\hat{i}_g} L_g({i}_g - i_g^\star)}_{\simeq\nabla\mathcal{S}} +~ \eta v_{dc} \,,    \label{eq: thetaint}
\end{align}\end{subequations}
where $K_f$ is a positive gain. 
A significant advantage of this approach is that one may limit the modulation magnitude by upper bounding the integrator \eqref{eq: gammaint} to e.g. $\mathrm{log}\tfrac{1}{\sqrt{2}}$.

As $\nabla\mathcal{S}$ acts directly on the coordinates $(\gamma_r,\theta)$, this synchronization gradient retains its physical interpretation as synchronization velocity (and orthogonal complement) and has been used in the experimental Section 5.5 of \cite{arghir2019transverse}, in a similar fashion, as a proxy to the main result. We refer the reader to Appendix \ref{appendix} for a further discussion on closed-loop stability.

Notice how the matching approach performs both DC-link regulation and current tracking, with just a single gradient-based controller of dimension 2, in this regard resembling the carrier-based Direct Power Control approach \cite{malinowski2004simple}. Furthermore, it does not require a PLL as, in the spirit of \cite{harnefors2021generic}, essentially incorporates one. 
As is the case in the cascaded control, grid impedance information is essential for gain tuning. By virtue of the OFO structure, one may also pursue a model-free implementation \cite{10354356}.

\begin{remark}\label{coupled_matching}
    Another aspect of this approach is the direct way in which the DC-link angle is coupled to the grid angle, in contrast to the cascaded PI structure, which introduces a delay in the coupling chain via the DC-bus regulator.
\end{remark}

\begin{figure}[!ht]
\centering{
\includegraphics[width=1\columnwidth]{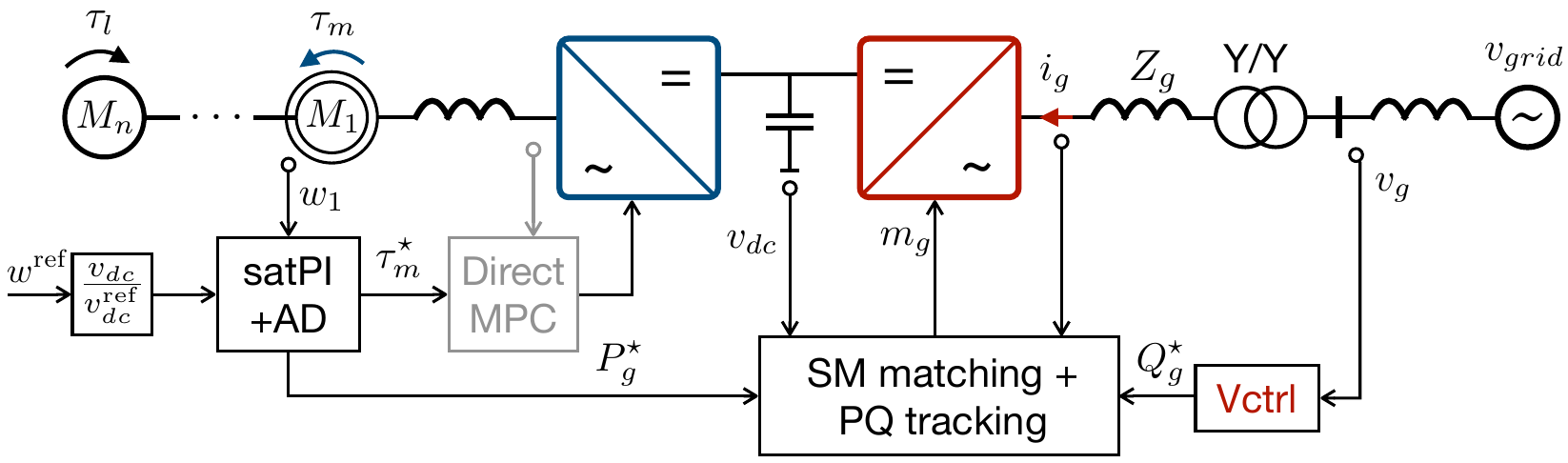}
\caption{Block diagram of the matching control setup.} 	\label{Fig: matchBlock}}
\end{figure}

For our experiments, we augment both controllers, illustrated in Fig. \ref{Fig: cascBlock} and \ref{Fig: matchBlock}, with a virtual impedance term, and subject them to weak and stiff grid conditions of SCR 1 and 49, respectively, with no change in the control gains throughout our tests. 

\section{Experimental results}
\label{expRes}

Through a proprietary software-in-the-loop setup, we simulate a pumped-hydro drive system with a 7MVA grid-side transformer 
and a 5kV DC-bus with mid-point. The converter is implemented using an average-model of the three-level active neutral-point-clamped (ANPC) topology. On the load-side, we implement a switching-model of the three-level ANPC converter connected to an Induction Machine (IM) model whose rotor is part of a 6MW driveline shaft modeled as a 5-mass torsional system as seen in Fig. \ref{Fig: dcmass}. 
The air-gap torque $\tau_m$ acts on one end of the shaft and the load $\tau_l$ on the other. 
%
The way in which the motor torque is produced (be it direct torque control, field-oriented control, or MPC) is not the focus of this work. 
For simplicity, we implement the projection $\Pi_\mathcal{C}$ via the activation-function approach in \cite{chandrasekaran2025reactive}.
For all our tests the speed reference is set to $w_\mathrm{nom}$ and we consider two grid scenarios: a stiff grid with an SCR of 49 and a weak grid with an SCR of 1. 
We have used $\kappa_\textit{pll} = 63$, while the matching control structure uses $K_f = {7.54\mathsf{e\!-\!}3}$. The cascaded control gains have been tuned with a significant separation of closed-loop bandwidth. For both approaches, we have used a virtual impedance $Z_v = 0.8 + J\omega_0{1\mathsf{e\!-\!}3}~\Omega$ added to the modulation vector as in Fig. \ref{Fig: matchDiag}.

{
While the theoretical treatment considered only one inductor element \eqref{igdynamics} between the converter and the measured voltage $v_g$, in our simulation we have used another (dynamic) inductor element between the measured voltage $v_g$ and the infinite bus $v_\textit{grid}$. These two elements resulted in }$Z_g = {1\mathsf{e\!-\!}2} + J\omega_0{9\mathsf{e\!-\!}4}~\Omega$, $Z_\textit{stiff} = {1.6\mathsf{e\!-\!}3} + J\omega_0{5.3\mathsf{e\!-\!}4}~\Omega$ in the stiff grid case, and $Z_\textit{weak} = {8\mathsf{e\!-\!}2} + J\omega_0{2.58\mathsf{e\!-\!}3}~\Omega$, i.e. approx. $49$ times larger, in the weak grid case.

\subsection{Phase jump test}

\begin{figure*}[hb]
\centering{
\includegraphics[width=1\textwidth]{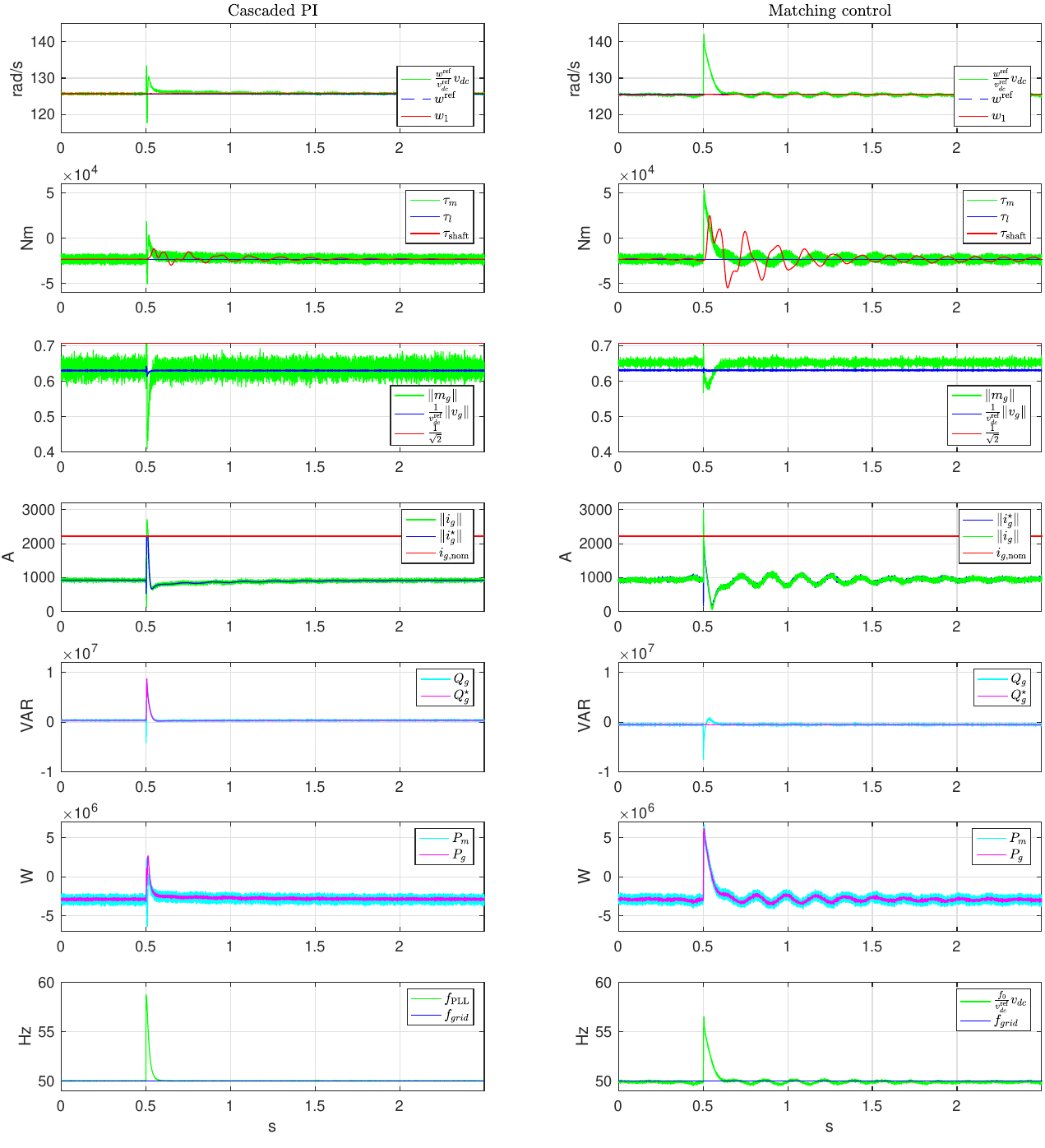}
\caption{Grid angle phase shift of 60 degrees, stiff grid response. We see that both the PLL and the matching voltage respond in a well damped manner. However, the matching control voltage has a larger excursion than the cascaded PI's $v_{dc}$ which incurs a larger response from the coupling control on the mechanical side and excites the TNF, as seen in the inner-shaft torque $\tau_\textit{shaft}$.}	\label{Fig: strong_Angle}}
\end{figure*}

\begin{figure*}[hb]
\centering{
\includegraphics[width=1\textwidth]{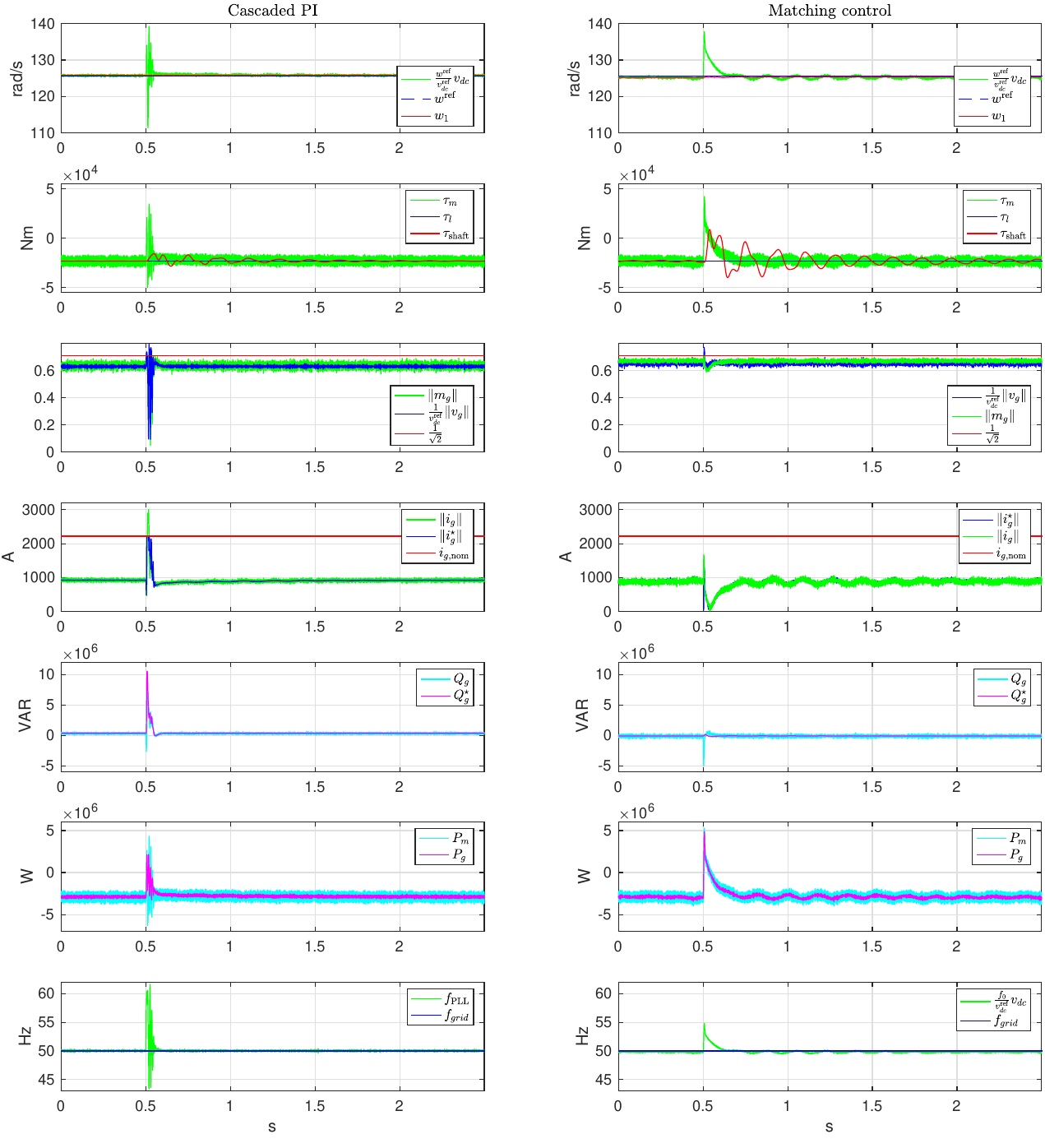}
\caption{Grid angle phase shift of 60 degrees, weak grid response. We see that the PLL has an underdamped response and cannot stabilize the AC voltage as well as the matching control, although it may be improved by tuning. This scenario shows the strength of the SM matching approach -- ride-through without reaching the current limit.}	\label{Fig: weak_Angle}}
\end{figure*}

We first set the load to $-0.5p.u.$, amounting to 2.93 MW of negative (generating) load. We compare side-by-side the response of the two control approaches to a 60-degree phase angle shift of the infinite bus $v_{grid}$. The stiff grid case is presented in Fig. \ref{Fig: strong_Angle} and the weak case in Fig. \ref{Fig: weak_Angle}.

\subsection{Three-phase drop test}

\begin{figure*}[hb]
\centering{
\includegraphics[width=1\textwidth]{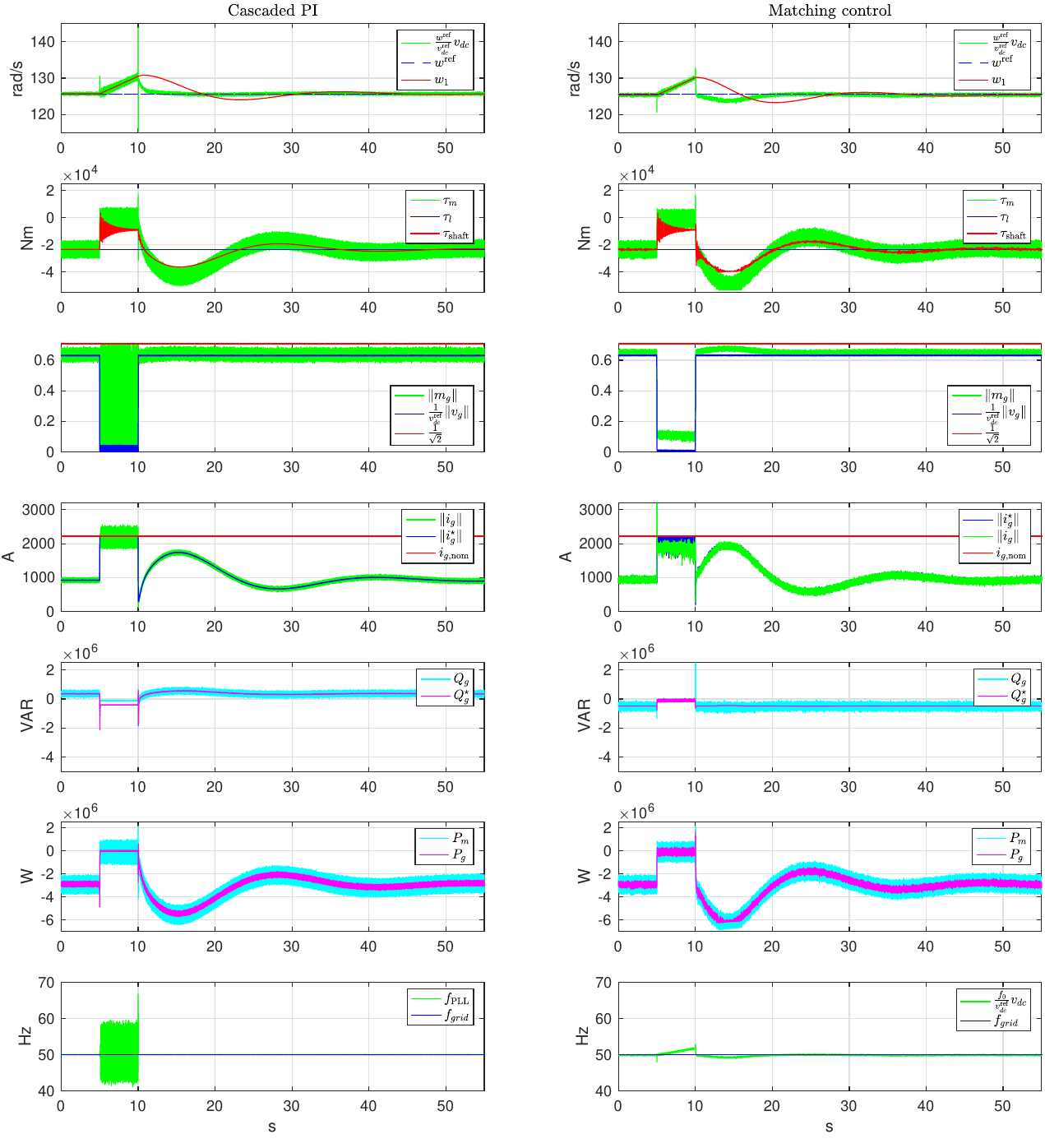}
\caption{Three-phase drop, stiff grid response. We see that the PCC drops to zero amplitude due to the low grid impedance. During this time the motor torque actually drops to zero preventing the DC-link to overcharge. This effect is simply the result of the elastic coupling that is enabled through the speed control and is beneficial. On the other hand, the reactive power is driven to zero due to the fact that the current magnitude saturates which forces the projection $\Pi_\mathcal{Q}$ to render zero reactive power demand. In fact, in the stiff grid situation, the AC voltage regulator has very little effect throughout the tests.}	\label{Fig: strong_Drop}}
\end{figure*}

\begin{figure*}[hb]
\centering{
\includegraphics[width=1\textwidth]{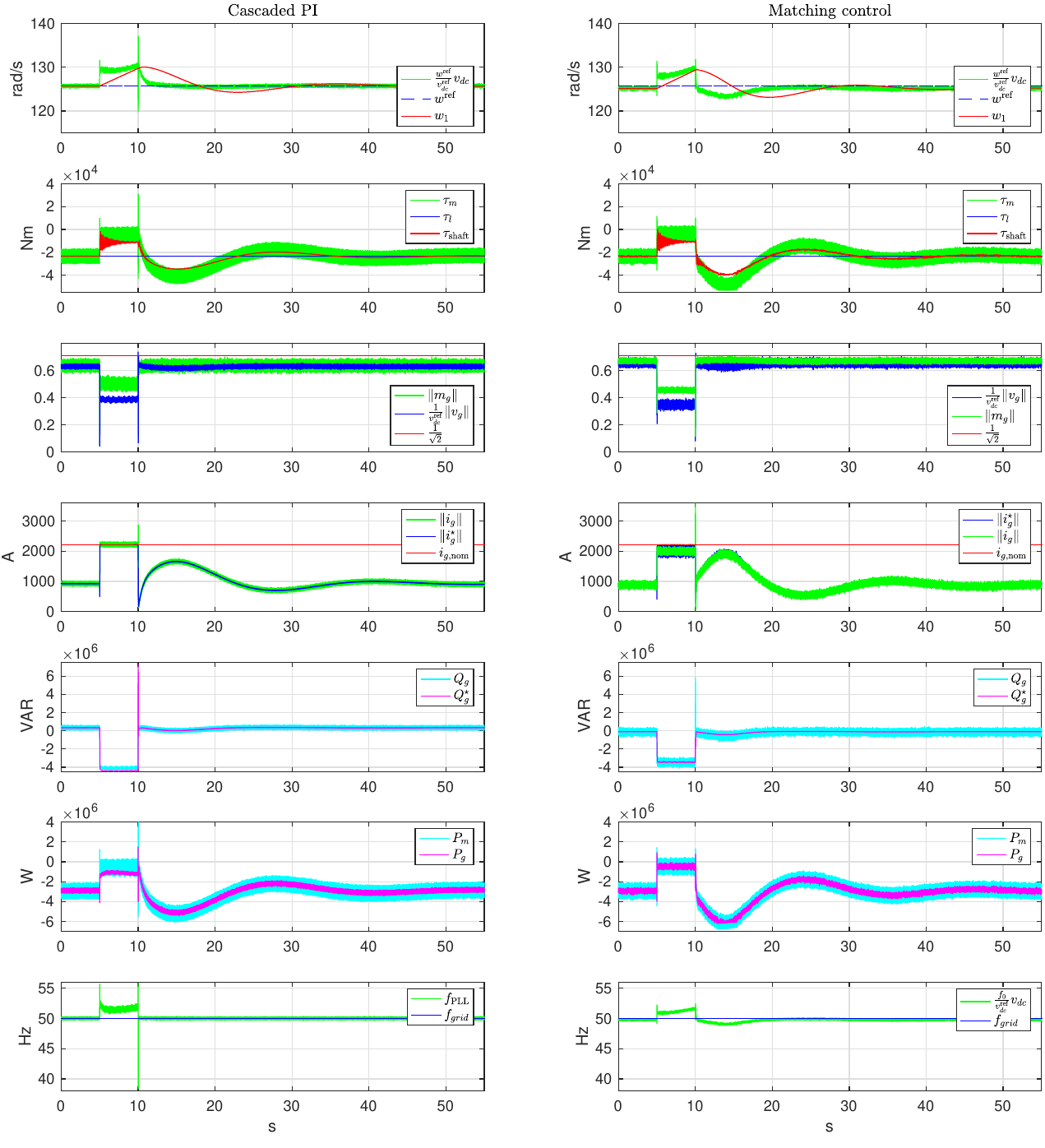}
\caption{Three-phase drop, weak grid response. Here we see that during the fault, the AC bus is sustained to the extent possible by a large capacitive current (negative $Q_g$) as a result of the AC voltage control \eqref{eq: qVC}. This feature is important for satisfying fault ride-through specifications found in e.g. IEC 61400. As before, $\tau_m$ drops to zero during the fault keeping the DC-link within limits. }	\label{Fig: weak_Drop}}
\end{figure*}

We then compare the response to a 5 second short-circuit of the infinite bus $v_{grid}$ at $-0.5p.u.$ mechanical load. The stiff and weak grid cases are presented in Fig. \ref{Fig: strong_Drop} and  \ref{Fig: weak_Drop}.

\subsection{Frequency step-up test}

\begin{figure*}[hb]
\centering{
\includegraphics[width=1\textwidth]{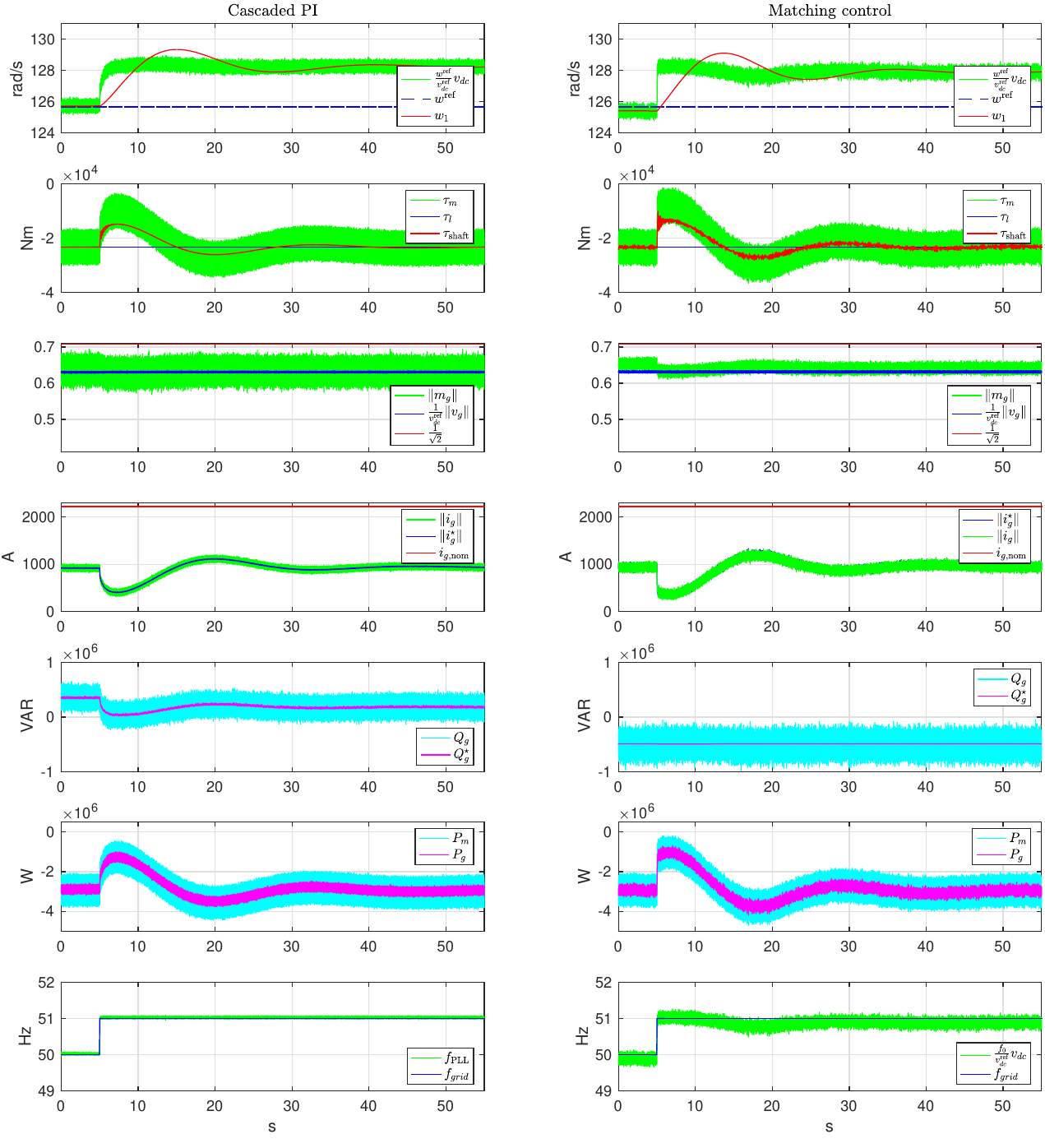}
\caption{Grid frequency step up, stiff grid response. In this test the infinite bus frequency is well tracked by both the PLL and the electronic synchronous machine. What is remarkable is how the velocity of the driveshaft tracks the DC-link causing an active power response to the frequency change, just as intended. The difference between the two approaches is noticeable in the DC-link transient, the cascaded PI is subject to the bandwidth of the DC-link control \eqref{vdc_PI} while the matching approach has an instant response.}	\label{Fig: strong_Fup}}
\end{figure*}

\begin{figure*}[hb]
\centering{
\includegraphics[width=1\textwidth]{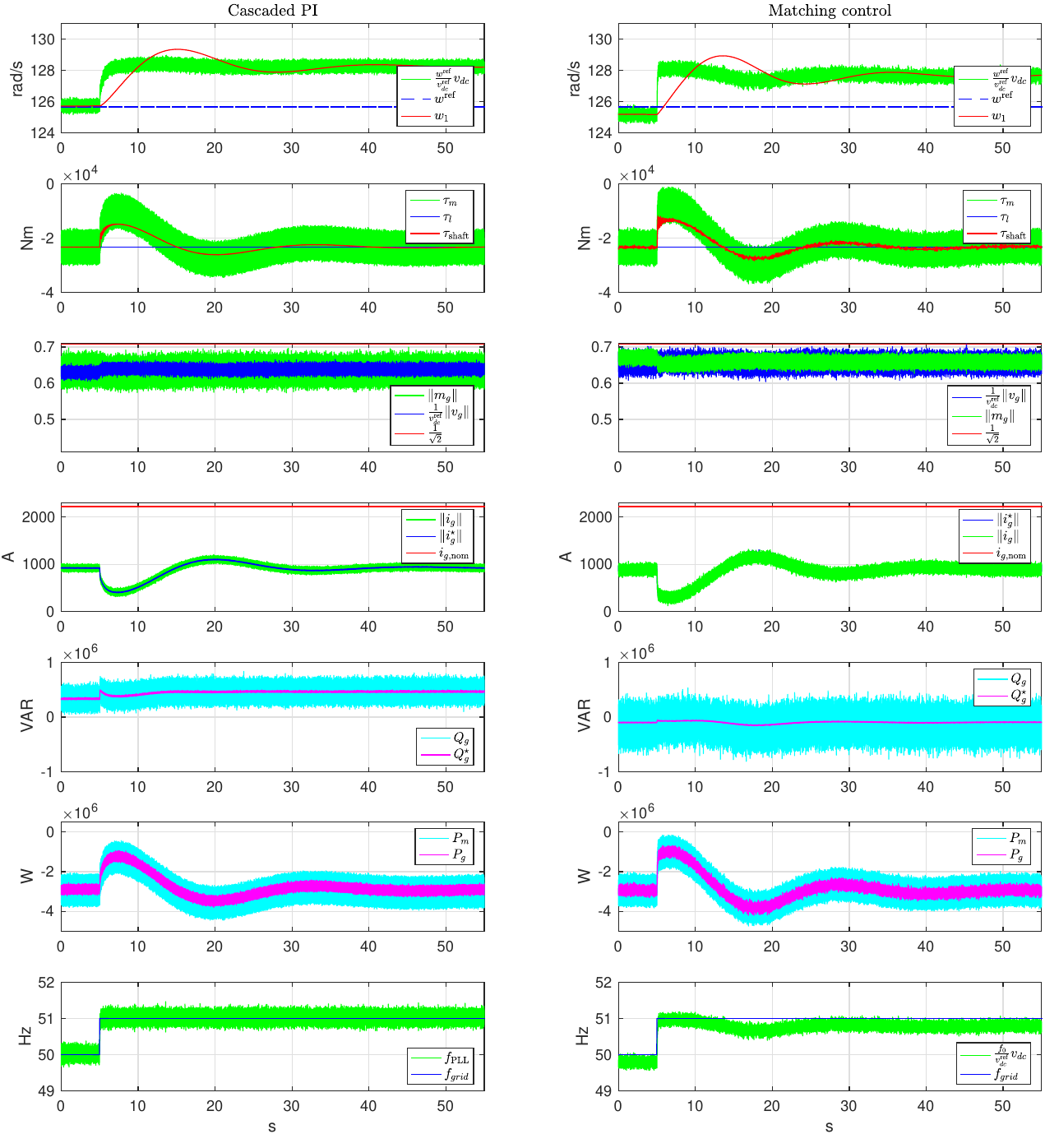}
\caption{Grid frequency step up, weak grid response. The difference here is mainly on the AC bus behavior, being more susceptible to disturbance and switching noise.}	\label{Fig: weak_Fup}}
\end{figure*}

Here we evaluate the response to a infinite bus $v_{grid}$ frequency step of $+1Hz$ again at $-0.5p.u.$. The stiff and weak grid cases are shown in Fig. \ref{Fig: strong_Fup} and \ref{Fig: weak_Fup}.

\subsection{Single-phase drop test}

\begin{figure*}[hb]
\centering{
\includegraphics[width=1\textwidth]{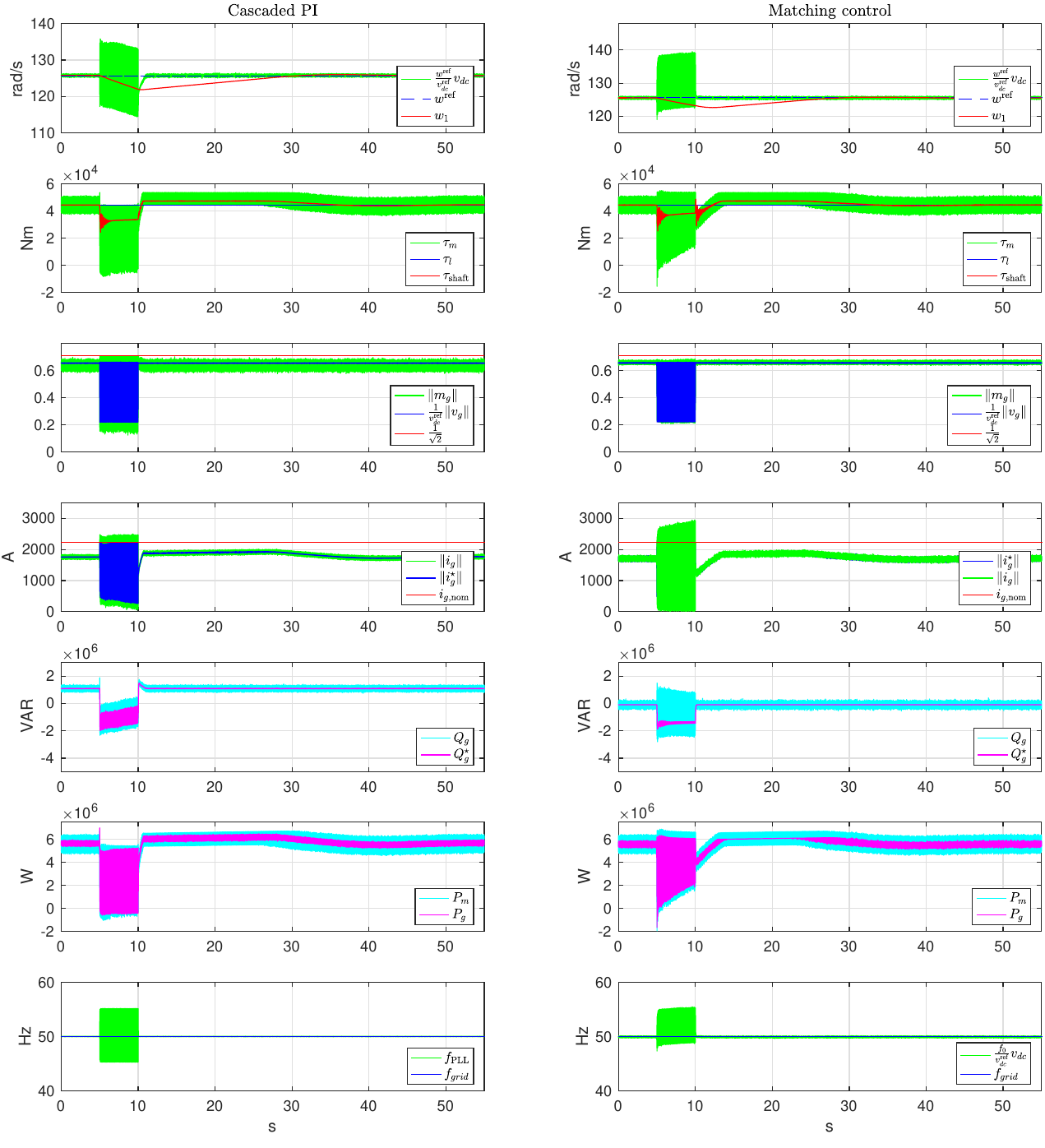}
\caption{Single phase fault, stiff grid response. Due to the low impedance of the stiff grid the second harmonic of the line frequency affects the drive in similar ways for both approaches.}	\label{Fig: strong_1ph}}
\end{figure*}

\begin{figure*}[hb]
\centering{
\includegraphics[width=1\textwidth]{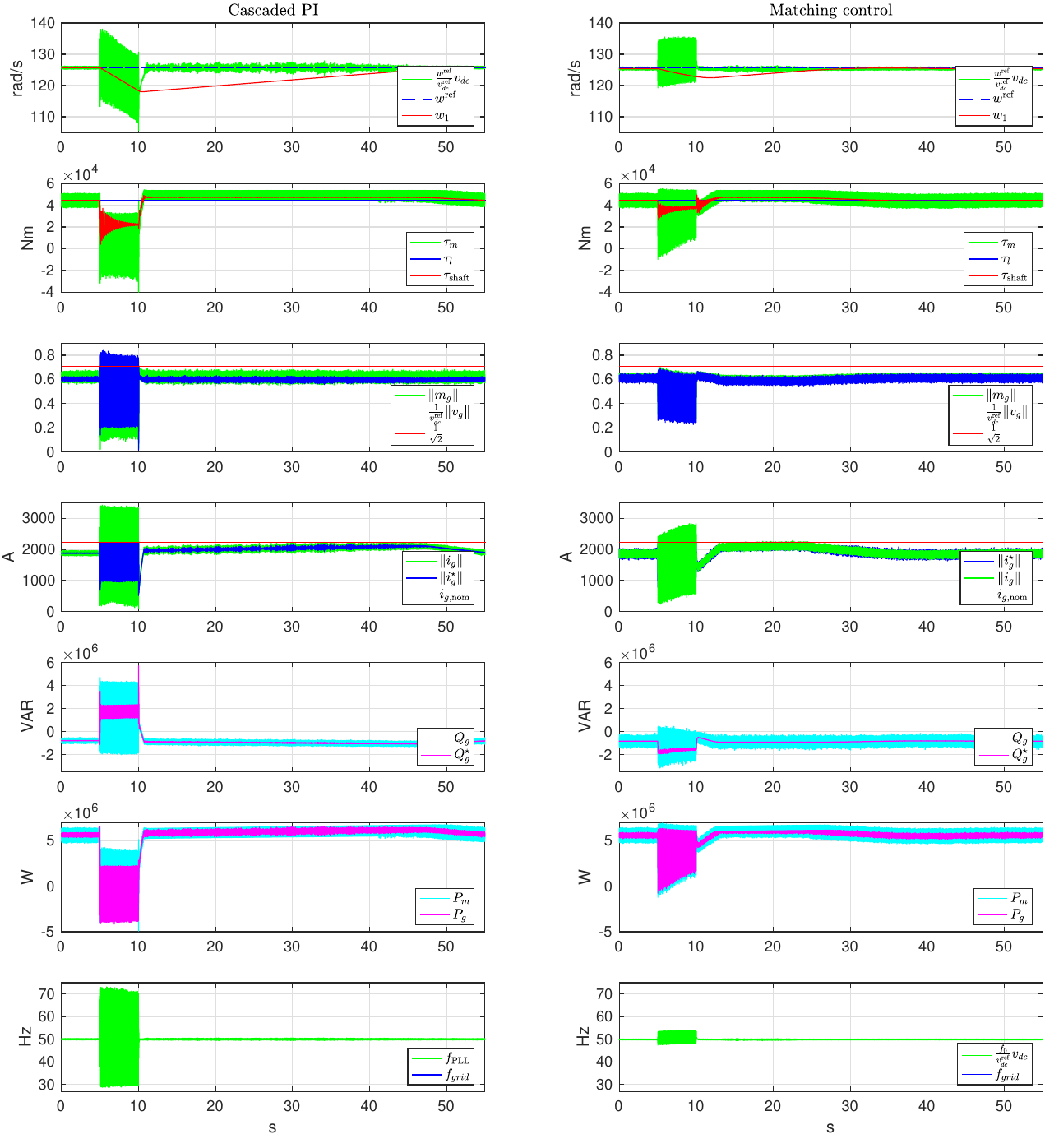}
\caption{Single phase fault, weak grid scenario. We see that the matching control has an advantage over the cascaded PI in reducing the effect of the second harmonic.}	\label{Fig: weak_1ph}}
\end{figure*}

We now set the load to $0.95p.u.$, amounting to about 5.57 MW of positive (motoring) load. We now compare side-by-side the response of the two control approaches to a drop of one phase to zero. The stiff grid case is presented in Fig. \ref{Fig: strong_1ph} and the weak case in Fig. \ref{Fig: weak_1ph}.

\subsection{Frequency step-down test}

\begin{figure*}[hb]
\centering{
\includegraphics[width=1\textwidth]{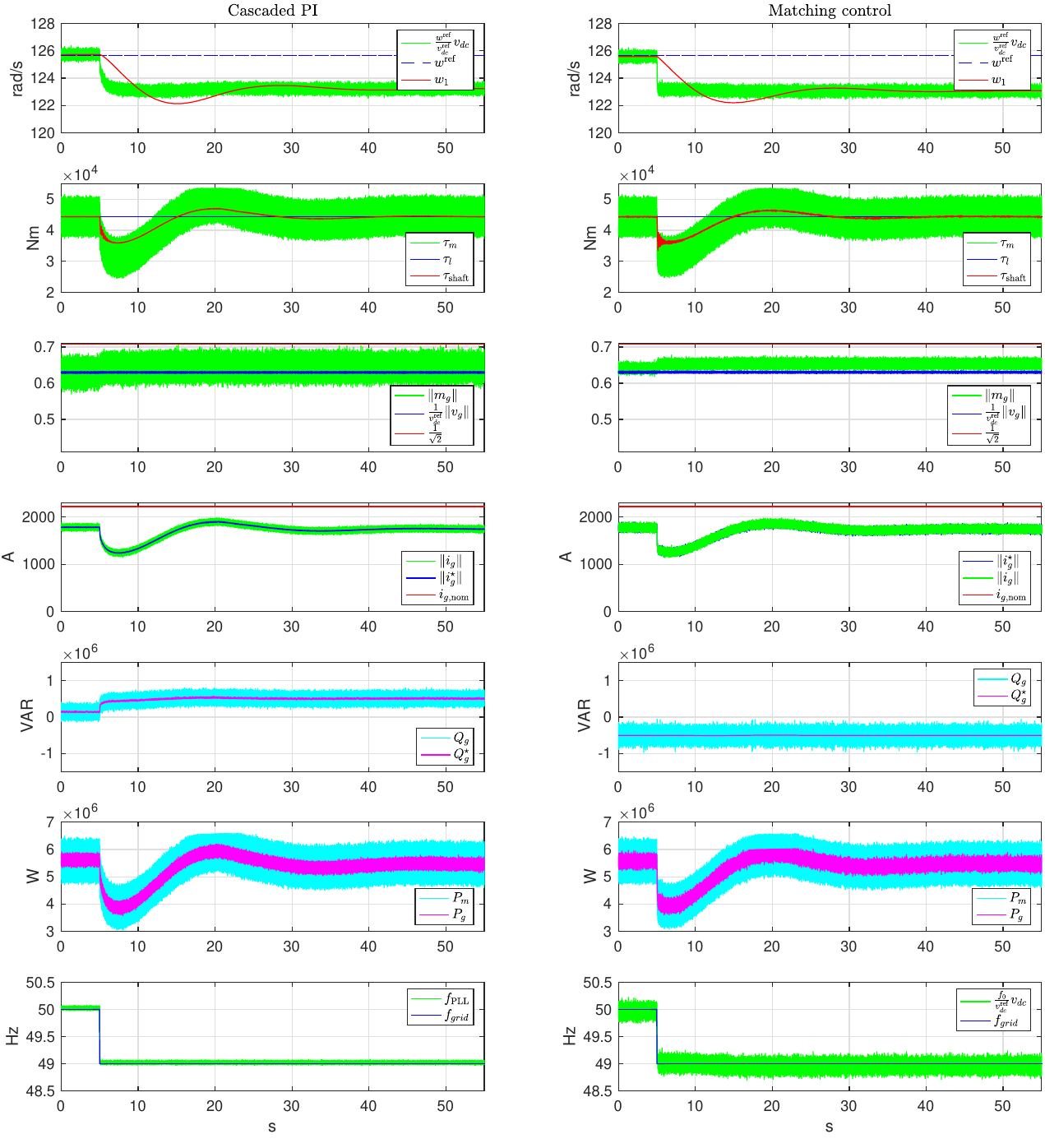}
\caption{Grid frequency step down, stiff grid response. In this case we see a very similar performance between the two approaches and when compared to Fig. \ref{Fig: weak_Fup}.}	\label{Fig: strong_Fdown}}
\end{figure*}

\begin{figure*}[hb]
\centering{
\includegraphics[width=1\textwidth]{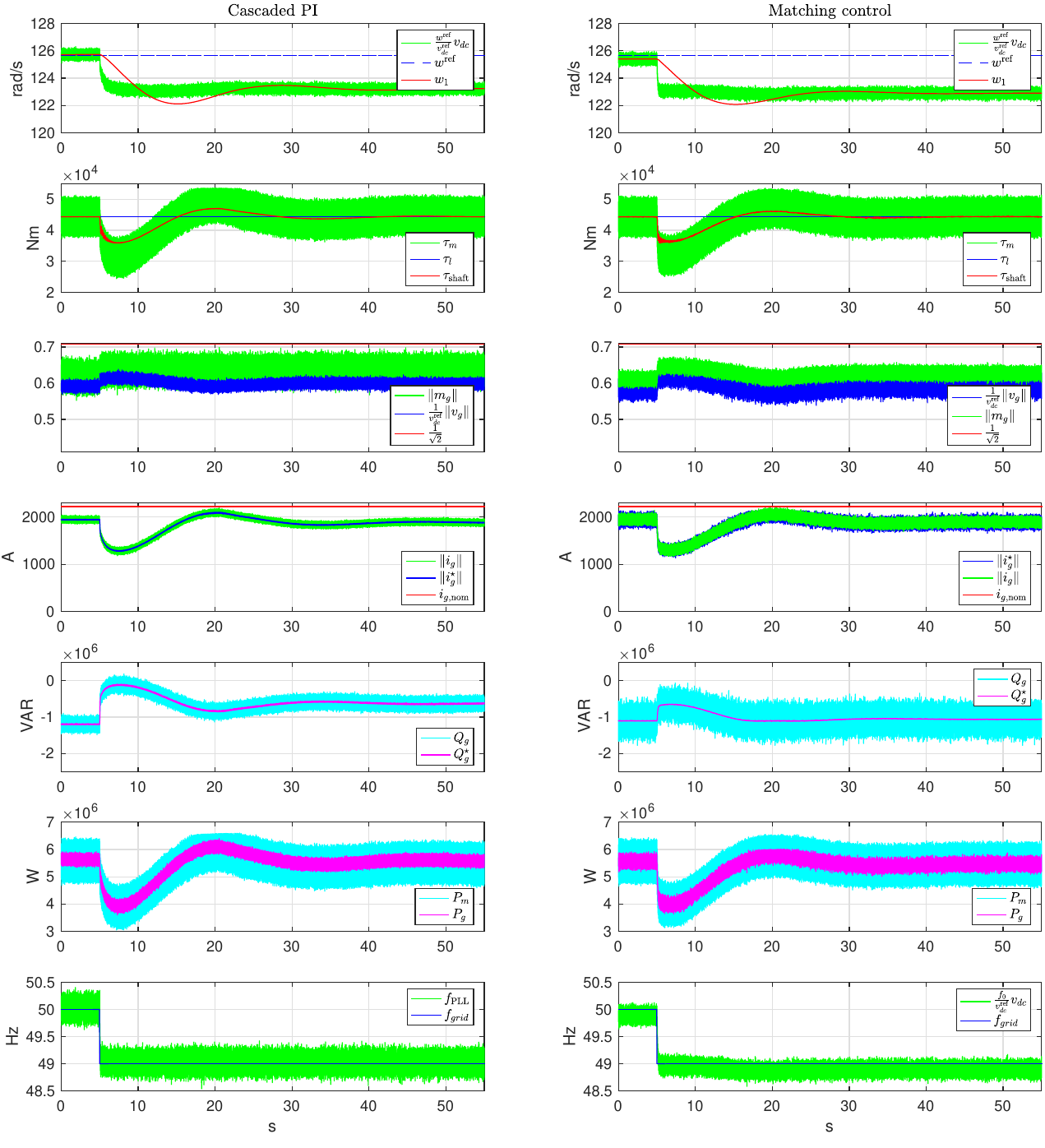}
\caption{Grid frequency step down, weak grid. As in Fig. \ref{Fig: weak_Fup}, the difference is mainly on the AC bus behavior, being more susceptible to disturbance and switching noise.}	\label{Fig: weak_Fdown}}
\end{figure*}

With the load at $0.95p.u.$, we evaluate the response to a infinite bus $v_{grid}$ frequency step of $-1Hz$. The stiff and weak grid cases are presented in Fig. \ref{Fig: strong_Fdown} and \ref{Fig: weak_Fdown}. Note that in both frequency tests, a step of $1Hz$ is an extreme case and represents a change in drivetrain power of approx. $0.02p.u.$, a difference which is seen at the end of the transient.

\subsection{Voltage dip test}

\begin{figure*}[hb]
\centering{
\includegraphics[width=1\textwidth]{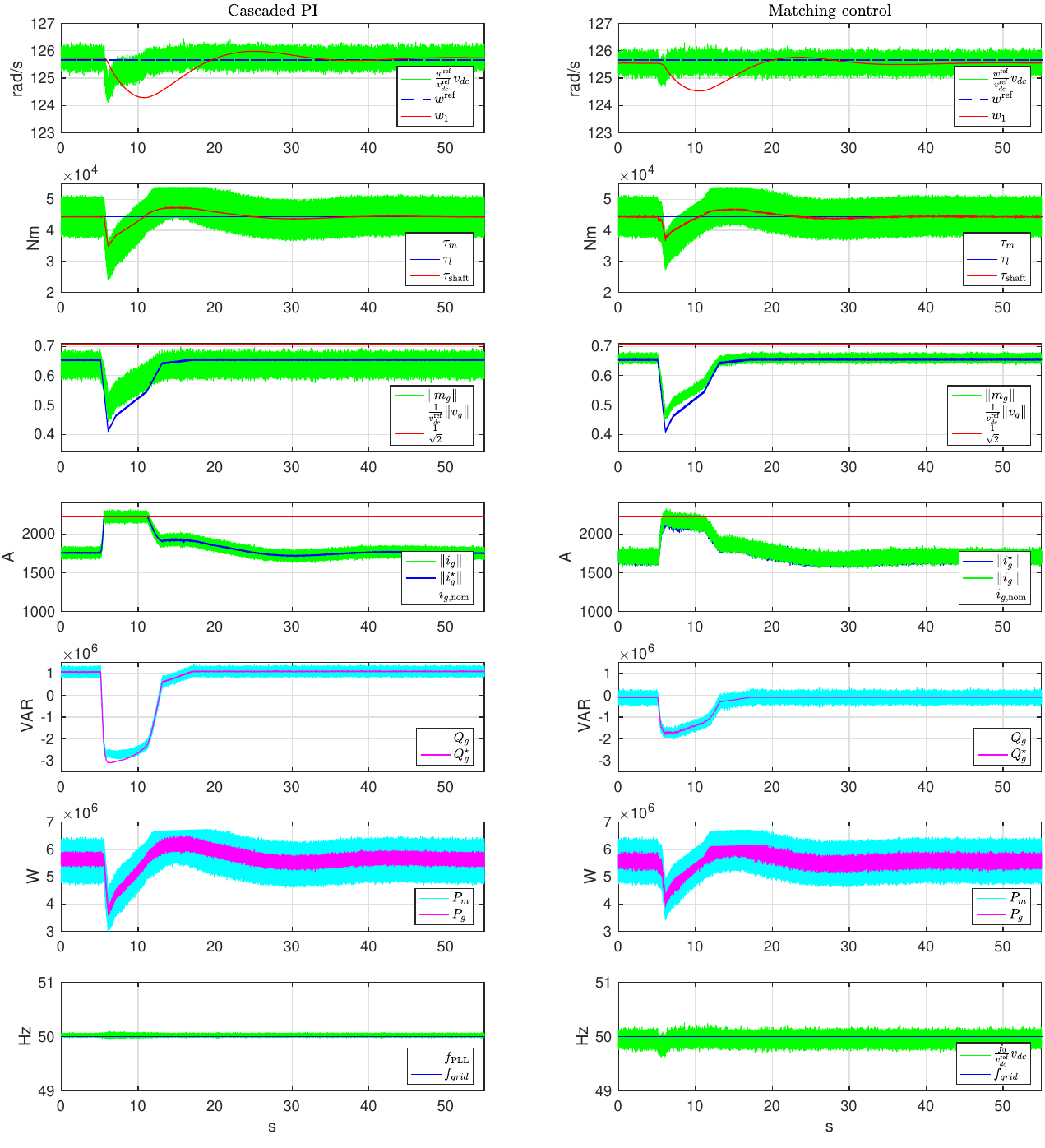}
\caption{Voltage dip, stiff grid scenario. As a direct comparison to \cite{chandrasekaran2025reactive}, we see a similar behavior here where the current is well limited through the reactive power controller as well as the barrier function mechanism which implements the function of projection onto the constraint set.}	\label{Fig: strongDip}}
\end{figure*}

\begin{figure*}[hb]
\centering{
\includegraphics[width=1\textwidth]{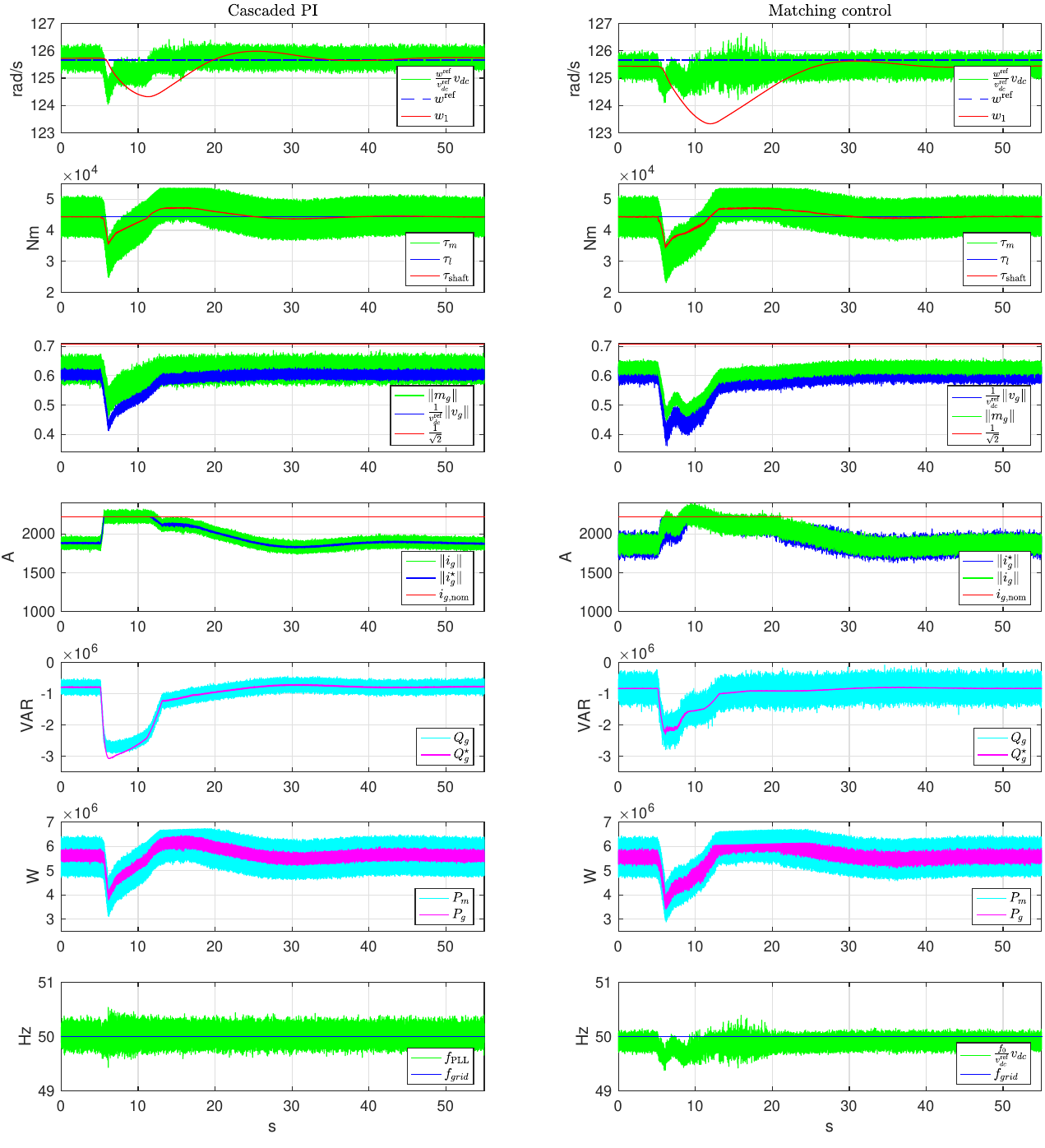}
\caption{Voltage dip, weak grid. We further note that the electronic synchronous machine behavior is different as the current tracking mechanism is different when compared to the cascaded PI. }	\label{Fig: weak_Dip}}
\end{figure*}

We now perform the test in \cite{chandrasekaran2025reactive}, Section VA under similar conditions (again $0.95p.u.$ load). The stiff grid case is presented in Fig. \ref{Fig: strongDip} and the weak case in Fig. \ref{Fig: weak_Dip}.

\subsection{Load reversal test}

\begin{figure*}[hb]
\centering{
\includegraphics[width=1\textwidth]{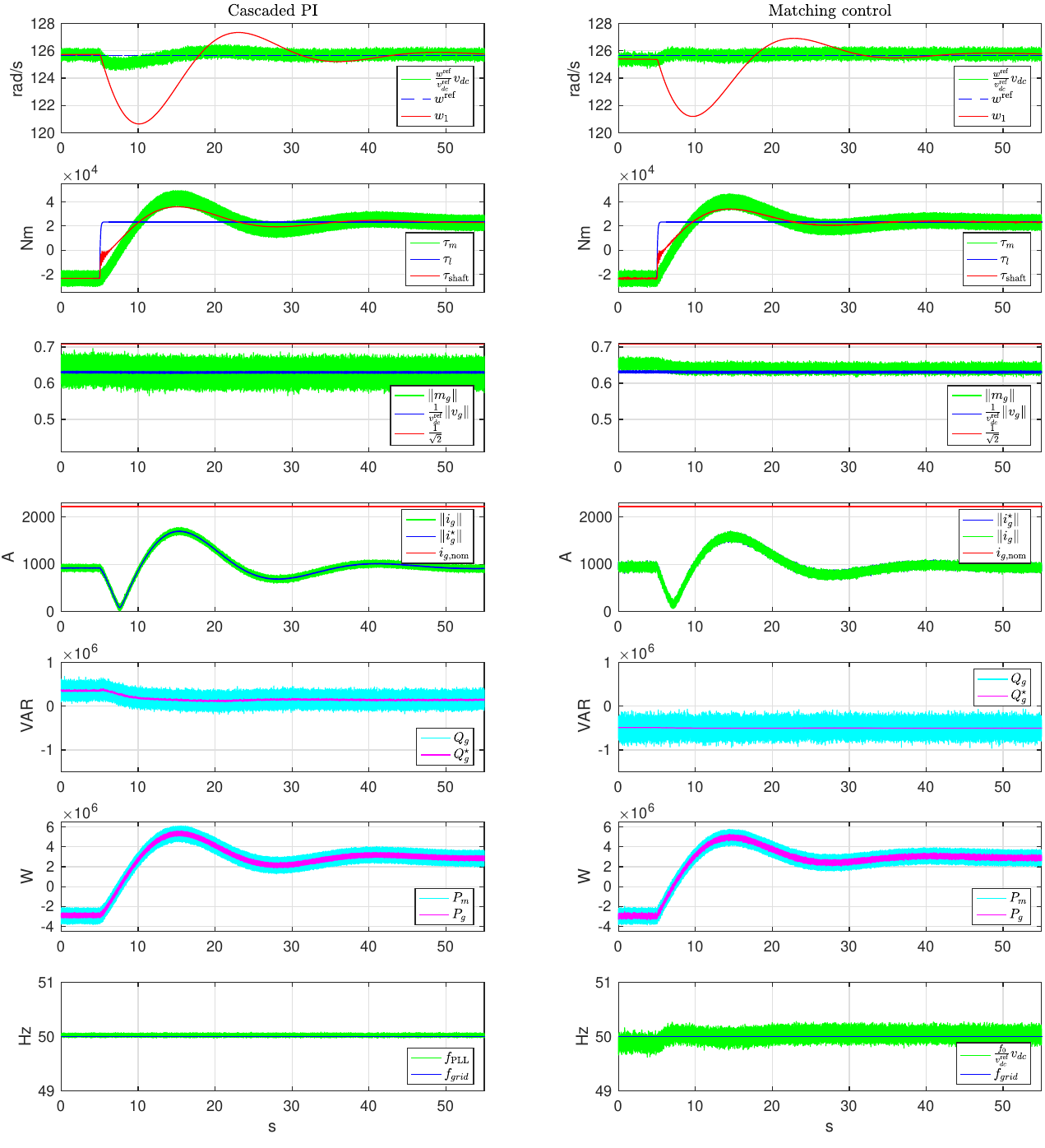}
\caption{Stiff grid, load step. The two control approaches yield similar results despite the difference in design philosophy.}	\label{Fig: strongStep}}
\end{figure*}

\begin{figure*}[hb]
\centering{
\includegraphics[width=1\textwidth]{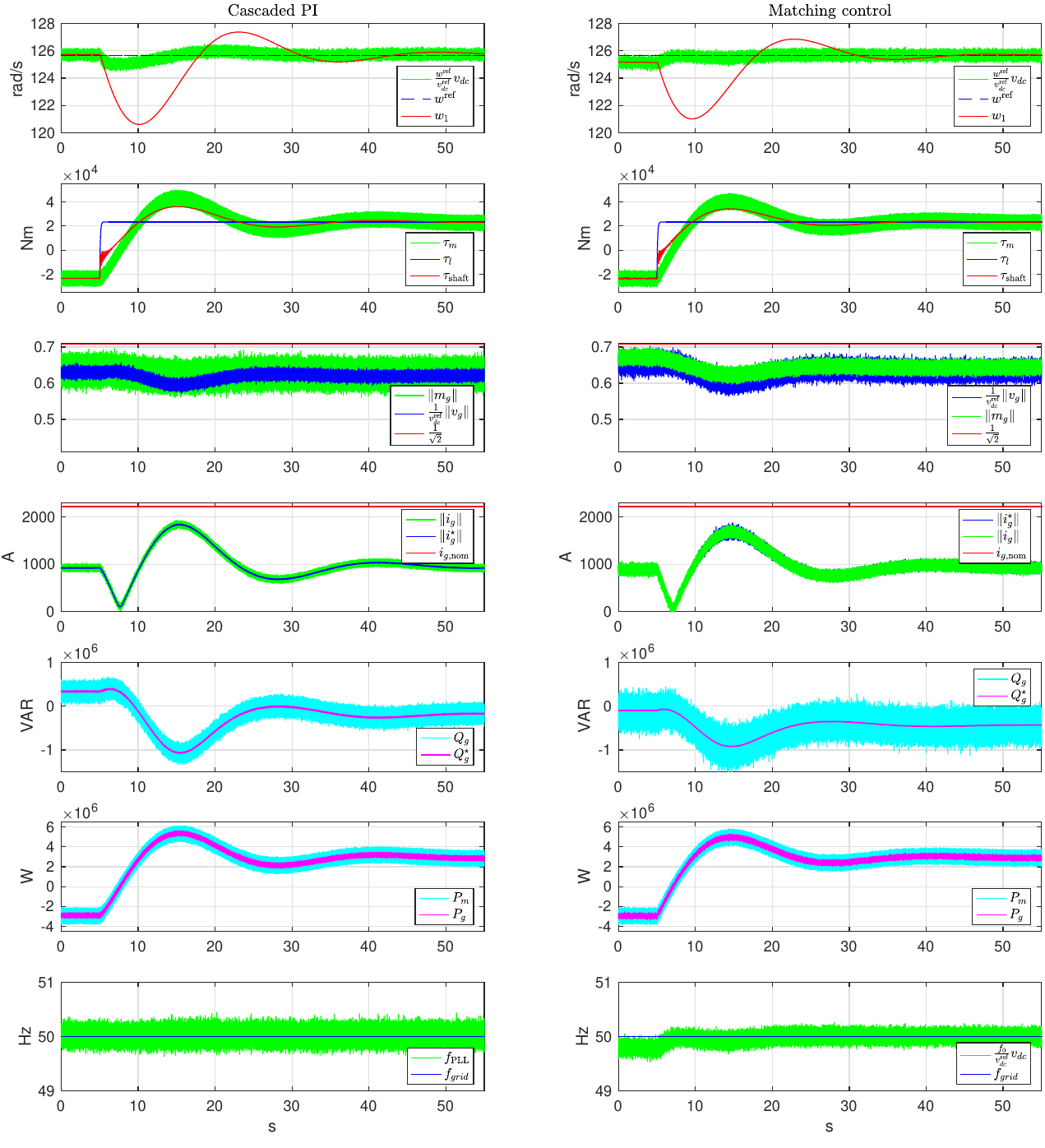}
\caption{Weak grid, load step. This test concludes the detailed comparison between the traditional cascade tracking approach and the novel, synchronous machine matching with energy-based control approach, yielding similar behavior. We see how the latter achieves current tracking, as well as modulation and current constraint satisfaction, in addition to the DC-link regulation, AC voltage support, and inertia provision features.}	\label{Fig: weak_Step}}
\end{figure*}

Finally, we present a load step from $-0.5p.u.$ to $0.95p.u.$ under conditions of stiff grid in Fig. \ref{Fig: strongStep}, and weak grid in Fig. \ref{Fig: weak_Step}.

\section{Conclusion}
\label{Concl}

We have proposed a two-step process to organically transfer the inertia of a drive system from the driveshaft to the grid by elastically coupling the driveshaft to the DC-bus and then, using a particular PLL mechanism, to couple the DC-bus to the grid frequency. In addition, we have proposed a novel approach to PLL design, where the angle and log-magnitude are used to minimize the Euclidean distance between the PLL output and the tracked voltage via gradient descent. We have further illustrated how this novel PLL design enables the standard matching control to incorporate a similar gradient descent mechanism that enables the tracking of a current set-point through its steady-state map. We therefore provide a framework for the matching control to completely replace the cascaded-control structure, without requiring a separate PLL mechanism, without an outer loop for DC-bus, and without suffering from the instabilities arising in weak grids, while preserving the same current limiting and tracking capabilities. One feature of the matching control method, most evident in the frequency-step tests, is that, when compared to the cascaded control approach, the frequency-to-power response is faster and it appears to have a higher effective DC-link regulation bandwidth. Another major significance of the flexible coupling approach is a natural, passivity-based approach to the interconnected structure of the driveline shaft, particularly in cases of complex shaft topologies where multiple motors are used.






\section{Appendix}\label{appendix}

As the DC-link and driveshaft dynamics are coupled have high inertia, we can verify, via a time-scale separation argument (i.e. when $v_{dc} = \textit{constant}$), the closed-loop stability on the grid side for both proposed methods. 

\subsection{A stability proof for the cascaded approach}\label{appendix0}

We first consider the cascade between the PLL system \eqref{eq: ePLL} and the grid-current dynamics \eqref{igdynamics} in closed loop with controller \eqref{ig_PI} (saturation disabled), while $v_g$ is at steady-state. We first write \eqref{eq: ePLL} by using the cartesian coordinate $v_\textit{pll}$ as
\begin{align}
    \dot{v}_\textit{pll} = -K_\textit{pll} e^{\gamma_\textit{pll}} ({v}_\textit{pll} - v_g) \,. \label{vPLL2}
\end{align}
Note that it achieves the steady-state characterized by $\theta_\textit{pll} = \arctan\tfrac{v_{g,\beta}}{v_{g,\alpha}}$, $\gamma_\textit{pll} = \log \|v_g\|$ and ${v}_\textit{pll} = v_g$. We now consider the cascade where \eqref{vPLL2} drives the subsystem \eqref{igdynamics},\eqref{ig_PI}, and where the PLL output is used in the controller. Let 
$$i_{g,dq} = R_{\theta_\textit{pll}}^\top i_g\,,$$ 
and $i_{g,dq}^* = \tfrac{1}{\|v_g\|}(P_g^*\mathrm{g}_1+Q_g^* J\mathrm{g}_1)$, where $P_g^*, Q_g^*$ are constant. Then, \eqref{igdynamics}, \eqref{ig_PI} are rewritten as
\begin{align}
    L_g\dot{i}_{g,dq} =& - R_g {i}_{g,dq} + v_g  - K_{p,g}({i}_{g,dq} -{i}_{g,dq}^*) - K_{i,g}x_g  \notag
    \\
    & - \dot{\theta}_\textit{pll} J L_g {i}_{g,dq} - {v}_\textit{pll} + Z_g{i}_{g,dq}^* \label{igdq_tilde}
    \\
    \dot{x}_g =&~ {i}_{g,dq} -{i}_{g,dq}^* \,. \notag
\end{align}
When system \label{vPLL2} is at steady-state, system \eqref{igdq_tilde} becomes
\begin{align}
    L_g\dot{\tilde{i}}_{g,dq} &= - Z_g \tilde{i}_{g,dq} - K_{p,g}\tilde{i}_{g,dq} - K_{i,g}x_g  \notag
    \\
    \dot{x}_g &= \tilde{i}_{g,dq}\,.
\end{align}
Using $\tfrac{1}{2} \tilde{i}_g^\top L_g \tilde{i}_g + \tfrac{1}{2} x_g^\top K_{i,g} x_g$ as Lyapunov function, it is easy to check that the driven system is globally asymptotically stable given when the driving system is at steady-state. Under the assumption of bounded trajectories, we can conclude that the cascade admits a globally asymptotically stable equilibrium $\{ {v}_\textit{pll,dq} = v_{g,dq}, {i}_{g,dq} = {i}_{g,dq}^*, x_g = 0 \}$.

\subsection{A stability proof for the matching approach}\label{appendix1}

We now consider again a cascade, this time between the $\theta,\gamma_r$ subsystem (embedded in the cartesian plane via $\hat{i}_g$) and ${i}_g$. If we take the first subsystem as
\begin{subequations}
\label{matching_law2}
\begin{align} 
    \dot\gamma_r &= - K_f \nabla_{\gamma_r}{\hat{i}_g} L_g(\hat{i}_g - i_g^\star) \label{eq: gammaint2}
    \\
    \dot\theta &= - K_f  \nabla_{\theta}{\hat{i}_g} L_g(\hat{i}_g - i_g^\star) + \omega_0 \,,    \label{eq: thetaint2}
\end{align}\end{subequations}
Let $\dot{\tilde\theta} = \dot\theta - \omega_0$ be the angle in $dq$-frame and define the euclidean embedding $\hat{i}_{g,dq} = - Z_g^{-1}\big( \tfrac{1}{\eta}e^{\gamma_r}{R}_{\tilde{\theta}}\mathrm{g}_1\omega_0 - v_{g,dq} \big)$. Then, \eqref{matching_law2} becomes
\begin{align} 
     \dot{\hat{i}}_{g,dq} = - \tfrac{e^{2\gamma_r}K_f \omega_0^2L_g}{\eta^2\|Z_g\|^2}(\hat{i}_{g,dq} - i_{g,dq}^*) \label{eq: gammadq}
\end{align}
which is asymptotically stable at $\{\hat{i}_{g,dq} = i_{g,dq}^*\}$. This can be seen as driving the subsystem defined by \eqref{igdynamics}, \label{eq: modlaw} when rotated by the grid angle. This becomes
\begin{align}
    L_g\dot{i}_{g,dq} =& - Z_g({i}_{g,dq} - \hat{i}_{g,dq}) \,,
\end{align}
which is globally asymptotically stable relative to the set $\{\hat{i}_{g,dq} = i_{g,dq}^*\}$. One can see how exploiting the cascade structure of the closed-loop systems, under both controllers, we are able to use LaSalle-type arguments or Proposition 14 in \cite{el2013reduction}, to conclude global asymptotic stability of the set-point $\{{i}_{g,dq} = i_{g,dq}^*\}$ for the grid subsystem under both control methods proposed. 

The complete stability proof, relying e.g. on time-scale separation assumptions, is not pursued here.

\subsection{Frequency domain analysis}\label{appendix2}

In order to compare the contribution of the driveline shaft inertia with the inertia given by the DC-link capacitor (when using matching control) we consider system \eqref{system_Mdc} where all quantities except for the current of the grid are expressed as the angular velocity (or angle). For obtaining the blue and green plots, we replace the term $\big(\tfrac{v_{dc}^\mathrm{ref}}{w^\mathrm{ref}}\big) m_g^\top i_g$ by an input $u$ and we linearize around an equilibrium where $w_{dc} = w_1 \neq 0$. In essence, the fraction $\tfrac{w_{dc}}{w_1}$ in \eqref{omega1dynamics_c} becomes equal to 1. We then plot the transfer function from input $u$ to output $w_{dc}$ in the coupled ($K_{p,m}, K_{i,m} > 0$) and uncoupled ($K_{p,m} = K_{i,m} = 0$) case. For obtaining the red and black plots we consider the transfer function between $\tau_m$ and $w_1$ for system \eqref{system_eq}. Note that the 2-mass system model is a (model-order) reduced version of a 5-mass system which is derived from a physical system.

\bibliographystyle{IEEEtran}
\bibliography{Inertia_control_TCST}

\end{document}